\documentclass[aps, pre, twocolumn, showpacs, superscriptaddress, floatfix]{revtex4-1}
\usepackage{amsmath,amsfonts, amsthm}
\usepackage{bm}
\usepackage{mathrsfs}
\usepackage{graphicx, subfigure}
\usepackage{subfigure}
\usepackage{verbatim}
\usepackage{mathrsfs}
\usepackage{hyperref}
\hypersetup{
        colorlinks=true,
}

\begin{document}

\newcommand{\up}{\uparrow}
\newcommand{\dn}{\downarrow}
\newcommand{\odiff}[2]{\frac{\di #1}{\di #2}}
\newcommand{\pdiff}[2]{\frac{\partial #1}{\partial #2}}
\newcommand{\di}{\mathrm{d}}
\newcommand{\ii}{\mathrm{i}}
\newcommand{\ua}{\uparrow}
\newcommand{\da}{\downarrow}
\renewcommand{\vec}[1]{{\mathbf #1}}
\newcommand{\vx}{{\bm x}}
\newcommand{\ket}[1]{|#1\rangle}
\newcommand{\bra}[1]{\langle#1|}
\newcommand{\pd}[2]{\langle#1|#2\rangle}
\newcommand{\tpd}[3]{\langle#1|#2|#3\rangle}
\renewcommand{\vr}{{\vec{r}}}
\newcommand{\vk}{{\vec{k}}}
\renewcommand{\ol}[1]{\overline{#1}}
\newtheorem{theorem}{Theorem}
\newcommand{\comments}[1]{}
\newcommand{\mysection}[1]{ \section{#1}}

\title{Fractional Josephson Effect in Number-Conserving Systems}
\author{Meng Cheng}
\author{Roman Lutchyn}
\affiliation{Station Q, Microsoft Research, Santa Barbara, California 93106-6105, USA}
\date{\today}
\begin{abstract}
We study fractional Josephson effect in a particle-number conserving system consisting of a quasi-one-dimensional superconductor coupled to a nanowire or an edge carrying $e/m$ fractional charge excitations with $m$ being an odd integer. We show that, due to the topological ground-state degeneracy in the system, the periodicity of the supercurrent on magnetic flux through the superconducting loop is non-trivial which provides a possibility to detect topological phases of matter by the {\it dc} supercurrent measurement. Using a microscopic model for the nanowire and quasi-one-dimensional superconductor, we derived an effective low-energy theory for the system which takes into account effects of quantum phase fluctuations. We discuss the stability of the fractional Josephson effect with respect to the quantum phase slips in a mesoscopic superconducting ring with a finite charging energy.
\end{abstract}

\pacs{
74.50.+r, 
73.21.Hb, 
71.10.Pm, 
74.78.Fk   
}

\date{\today}

\maketitle

\section{Introduction}
Josephson effect, a hallmark of the macroscopic quantum coherence, has played a crucial role in the research and applications of superconductivity, both conceptually and practically since its discovery~\cite{JosephsonBD, Golubov_RMP'04, Buzdin'05, Bergeret'05, KulikReview}. Recently, it has been recognized that Josephson effect can also reveal the topological aspects of superconductivity, most notably the $4\pi$-periodic ac Josephson effect~\cite{1DwiresKitaev, Kwon} in one-dimensional class D topological superconductors~\cite{1DwiresKitaev, Kwon, Kwon2, MajoranaQSHedge, 1DwiresLutchyn, 1DwiresOreg}, as well as in time-reversal-invariant generalizations~\cite{ChungPRB2013, LiuPRX2014}. The doubling of the periodicity is tied to the existence of non-trivial excitations - localized Majorana zero-energy modes (MZMs) at the opposite ends of a topological superconductor. The presence of the zero-modes allows for the coherent charge $e$ tunneling processes between two superconductors. This, in turn, leads to the doubling of the flux periodicity as compared to the conventional Josephson effect involving charge $2e$ (Cooper-pair) tunneling.

In the last five years there has been a surge of research interest in topological superconductors~\cite{Fu:2008, MajoranaQSHedge, Sau, Alicea, 1DwiresLutchyn, 1DwiresOreg, CookPRB'11, SauNature'12, Mourik2012, Rokhinson2012, Das2012, Deng2012, Fink2012, Churchill2013, Chang_PRL2013, Lee_arxiv2013, Deng_arxiv2014, Nadj-Perge2014}. Apart from the fundamental importance, the search for Majorana zero-modes and other non-Abelian quasiparticles is fueled by the prospects of topological quantum information processing~\cite{TQCreview, AliceaBraiding, SauWireNetwork, ClarkeBraiding, TopologicalQuantumBus, BraidingWithoutTransport}. Recent theoretical breakthrough indicating that Majorana-based topological phases can be accessed in heterostructures involving a semiconductor nanowire coupled to a conventional s-wave superconductor~\cite{1DwiresLutchyn, 1DwiresOreg} has sparked a significant experimental activity on this subject~\cite{Mourik2012, Rokhinson2012, Das2012, Deng2012, Fink2012, Churchill2013, Chang_PRL2013, Lee_arxiv2013, Deng_arxiv2014, Krogstrup2014, Chang2014, Higginbotham2015}. Apart from Ref.~[\onlinecite{ Rokhinson2012}], most of the aforementioned experiments, however, have been focusing on the zero-bias peak anomaly associated with presence of the zero-energy modes~\cite{ZeroBiasAnomaly0,ZeroBiasAnomaly1,ZeroBiasAnomaly2,ZeroBiasAnomaly3,ZeroBiasAnomaly31, ZeroBiasAnomaly4,ZeroBiasAnomaly5,ZeroBiasAnomaly6, 1DwiresLutchyn2, ZeroBiasAnomaly61, ZeroBiasAnomaly7}.

\begin{figure}
	\centering
	\includegraphics[width=0.8\columnwidth]{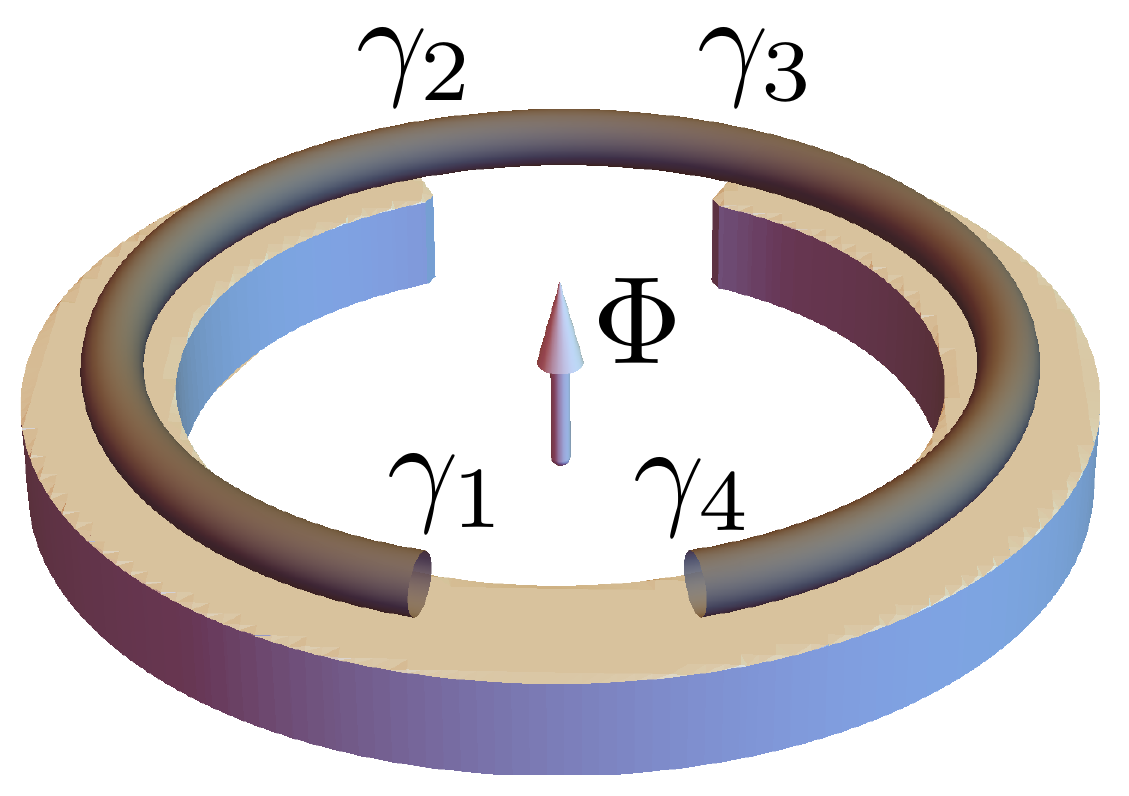}
	\caption{Proposed setup to study Josephson effect in particle-number conserving system. a) The nanowire is coupled to a quasi-one-dimensional superconductor forming a loop with the circumference $L$. The four Majorana zero modes $\gamma_{1,2,3,4}$ are located at $x=L-l-l', L-l', 0, l$ respectively. The distance between $\gamma_2$ and $\gamma_3$ is $l'$ and the distance between $\gamma_1$ and $\gamma_2$ is $l$, same as the distance between $\gamma_3$ and $\gamma_4$. A magnetic flux $f$ pierces through the loop and induces the supercurrent flow in the loop. The periodicity of the supercurrent on flux allows one to probe topological properties of the system. In the case of parafermion proposals, the Luttinger liquid corresponds to, for example, an edge of a two-dimensional Abelian fractional quantum Hall system.}
	\label{fig:setup}
\end{figure}

Fractional Josephson effect provides perhaps the simplest setup where the non-Abelian nature of the Majorana zero modes are manifested. The stability of this effect under various realistic situations (i.e. energy splitting, disorder, multiple bands, quasiparticle poisoning) have been extensively studied in literature~\cite{LawPRB2011, ChengPRB2012, PabloPRL2012, ZazunovPRB2012, Sau_unpub, Sau_unpub2, Fernando_PRB2012, PientkaNJP, Houzet_PRL2013, Doru_PRB2013, AldoPRB2013, Caldeira_unpub, Crepin}. The standard approach to understand the Josephson effect is based on BCS mean-field theory, where the $\mathrm{U}(1)$ particle number conservation is spontaneously broken. However, due to the mesoscopic nature of the experimental setups in engineering topological superconductors, quantum phase fluctuations may play an important role. Indeed, charging energy is at the heart of many mesoscopic superconducting devices such as superconducting qubits. Furthermore, charging energy is important for topological quantum computing schemes with Majorana zero modes~\cite{Flensberg_PRL2011, Martin'11, Wang'2011, Liu'11, TopologicalQuantumBus, Martin'2012, BraidingWithoutTransport, Hyart2013}. There have been several works that take into account quantum phase fluctuations in a phenomenological way~\cite{Fu_Coulomb, HeckPRB2011, PekkerPRB2013, MatthewsPRL2013, Houzet_PRL2013, Fernando_PRB2012}. However, a microscopic theory of fractional Josephson effect is still lacking. In this work, we start from a microscopic model of helical nanowires coupled to a fluctuating one-dimensional s-wave superconductor introduced in Ref. [\onlinecite{Fidkowski2011}]. Although there is no long-range superconducting order, there are still Majorana zero modes and a related topological degeneracy when two or more nanowires are coupled to the same quasi-one-dimensional superconductor (QSC)~\cite{Fidkowski2011, SauHalperin2011}. We study the Josephson effect within this model and find that fractional Josephson effect survives quantum fluctuations. However, the splitting of the ground state degeneracy as well as the hybridization between different topological sectors changes and now becomes power-law dependent on the system size. In order to show that we consider instanton tunneling events between different topological sectors. We find that phase slip events at weak links caused by the spatial inhomogeneities (e.g. impurities) in the quasi-one-dimensional superconductor are responsible for the power-law decay with length of the ground-state energy splitting. Finally, we calculate the periodicity of the Josephson current on magnetic flux through the ring, see Fig.~\ref{fig:setup} for the proposed setup. It has been previously believed that one needs to perform an ac measurement to detect $4\pi$-periodicity of the supercurrent~\cite{1DwiresKitaev, Kwon} which might be quite challenging due to parity switching processes (quasiparticle poisoning)\cite{MajoranaQSHedge, 1DwiresLutchyn} as well as Landau-Zener transitions to the quasiparticle continuum~\cite{Houzet_PRL2013, Badiane2013} which together put constraints on the appropriate time window for performing the experiment, see below. In this paper, we show that by suitably designing the experimental system, one can avoid the aforementioned problems and may be able to detect $4\pi$-periodicity in the {\it dc} experiments.

Our approach based on Luttinger liquid formalism allows one to extend our theory to the recently proposed systems hosting parafermionic zero modes~\cite{Clarke12, Lindner12, cheng2012}. In this case, we consider an edge of two-dimensional Abelian fractional quantum Hall system properly coupled to QSC. Our main conclusions discussed above for the Majorana case remain also valid for parafermions, and we show how to probe topological properties of parafermions in the dc measurements, see Fig.~\ref{fig:setup}.

The paper is organized as follows. We begin with the qualitative discussion section~\ref{sec:qual} where we explain our results using a simplified toy model which captures some main aspects of the fractional Josephson effect. This model, however, does not take into account superconducting quantum phase slips. In order to treat quantum fluctuations properly, one needs to develop a different formalism using Abelian bosonisation technique. For pedagogical reasons, we first review Josephson effect in a Luttinger liquid coupled to a bulk s-wave (Sec.~\ref{sec:swave}) and p-wave (Sec.~\ref{sec:pwave}) superconductors, and discuss the mechanisms for the change of a periodicity with magnetic flux in these two systems. Next, in Sec.~\ref{sec:parafermion} we review the fractional Josephson effect in the parafermion systems by considering an edge of a two-dimensional Abelian fractional quantum Hall system at the filling fraction $\nu=1/m$ coupled to a bulk superconductor. In Sec.~\ref{sec:number_cons}, we present our main results for Josephson effect in a number-conserving setup and discuss the dependence of the ground-state energy of the system on magnetic flux. Finally, the effect of quantum phase fluctuations on the flux periodicity of the Josephson current is discussed in Sec.~\ref{sec:splitting}. Some technical details are presented in the Appendix~\ref{app:splitting}.

\section{Qualitative Discussion of Main Results}\label{sec:qual}

First, we discuss fractional Josephson effect using the simple model involving bulk superconductor with the long-range order. Let us consider the setup shown in Fig.~\ref{fig:setup}. There are four Majorana zero modes $\gamma_{1,2,3,4}$ residing at the ends of the 1D topological superconductors, and the corresponding effective low-energy Hamiltonian reads
\begin{equation}
	H\!=\!iE_J\gamma_2\gamma_3\cos\frac{\Phi}{2}\!+\!i\delta E_{12}\gamma_1\gamma_2\!+\!i\delta E_{34}\gamma_3\gamma_4\!+\!i\delta E_{14}\gamma_1\gamma_4.
	\label{eqn:bcs}
\end{equation}
Here $E_J$ is the $4\pi$ Josephson coupling at the junction due to the hybridization between Majorana modes $\gamma_2$ and $\gamma_3$, $\delta E_{ij}$ are the energy splittings between Majorana modes $\gamma_i$ and $\gamma_j$ which are exponentially small with the distance between them $L_{ij}$: $\delta E_{ij}\propto \exp(-L_{ij}/\xi_{ij})$ with $\xi_{ij}$ being the superconducting coherence length in the corresponding segment. It is important to notice that $\delta E_{ij}$, in particular $\delta E_{14}$, are independent of the magnetic flux due to the large superfluid stiffness of the bulk superconductor.

\begin{figure}
	\centering
	\includegraphics[width=3in]{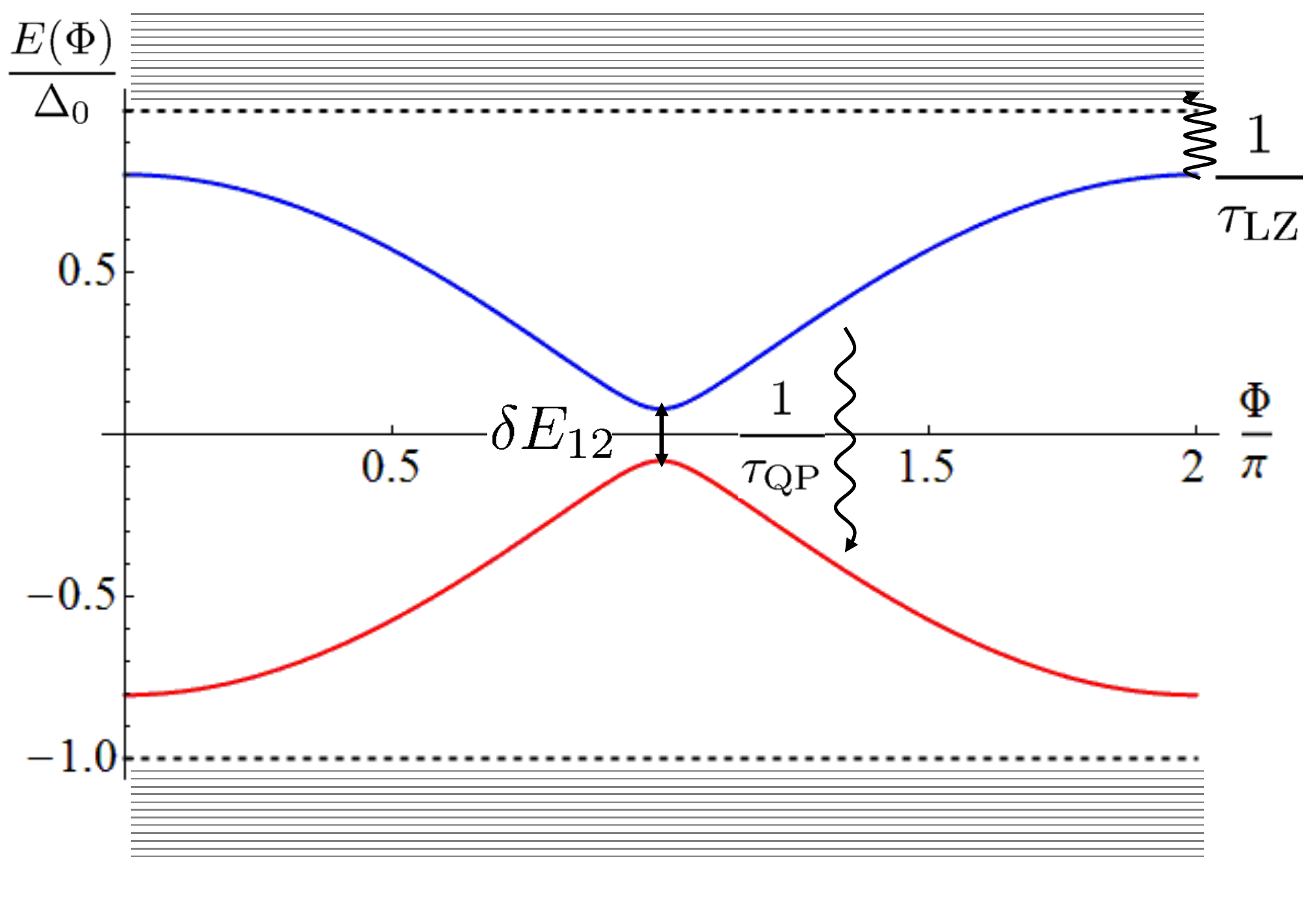}
	\caption{Fractional {\it ac} Josephson effect. The spectrum of the two Andreev bound states corresponding to different fermion parity is hybridized at $\Phi=\pi$ by the exponentially-small energy splitting $\delta E_{12}$. Diabatic passage through the avoided level-crossing allows one to follow fixed fermion parity state and to detect the $4\pi$ Josephson effect in {\it ac} measurements. The rate of the superconducting phase advance is constrained by the parity relaxation (quasiparticle poisoning) and Landau-Zener transition rates defined by $1/\tau_{\rm QP}$ and $1/\tau_{\rm LZ}$, respectively.}
	\label{fig:andreevspec0}
\end{figure}

 We now review the physics of the {\it ac} Josephson effect~\cite{1DwiresKitaev, Kwon}. For pedagogical reasons, it is useful to assume that $\delta E_{14}\rightarrow 0$ and $\delta E_{12}=\delta E_{34}$. In the thermodynamic limit $\delta E_{12}\rightarrow 0$, the spectrum of Andreev levels is given by $\pm E_J \cos(\Phi/2)$ where different states correspond to even/odd fermion parity of the modes at the junction. If the fermion parity is preserved over the time evolution of the superconducting phase $\Phi$ from $0$ to $2\pi$, one can see that the system ends up in the excited states, and thus the Josephson current through the junction will be $4\pi$ periodic. If we take into account finite-size effects, one can show that Andreev levels do not quite cross at $\Phi=\pi$ due to the exponentially small splitting energy $\delta E_{12}$, see Fig.~\ref{fig:andreevspec0}. Nevertheless, provided the evolution of the phase $\Phi(t)=\alpha t$ is fast enough (i.e. $\alpha \gg \delta E_{12}$), the fermion parity of Andreev levels will be approximately preserved, and one may still hope to detect $4\pi$ Josephson effect in {\it ac} measurements. This phenomenon was dubbed as Fractional ac Josephson effect.

It was shown recently~\cite{MajoranaQSHedge, 1DwiresLutchyn, Houzet_PRL2013, Badiane2013} that in order to understand the dynamics of the Andreev bound states, which is important for the experimental detection of the fractional ac Josephson effect, one also needs take into account relaxation processes due to quasiparticle poisoning as well as Landau-Zener transitions to the quasiparticle continuum, see Fig.~\ref{fig:andreevspec0}. The former corresponds to parity relaxation processes due to the presence of stray non-equilibrium quasiparticles whereas the latter represents the Landau-Zener transitions during the diabatic passage. Roughly speaking, the rate at which the phase $\Phi$ is ramped should be fast compared to the
splitting energy at the avoided crossing at $\Phi=\pi$, and fermion parity relaxation rate due to non-equilibrium quasiparticles. On the other hand, the sweep rate of $\Phi$ cannot be too fast, otherwise unwanted Landau-Zener transitions into continuum at, for example, $\Phi=2\pi$ would become significant. Another relevant timescale $\tau_R$ is associated with the superconducting phase relaxation dynamics due to the resistive environment in which the Josephson junction is embedded and, as such, depends on the specific experimental setup~\cite{Houzet_PRL2013, Badiane2013}. Overall, experimental observation of the fractional {\it ac} Josephson effect is quite challenging because one needs to know a priori the aforementioned timescales.

\begin{figure}[htpb]
	\centering
	\includegraphics[width=\columnwidth]{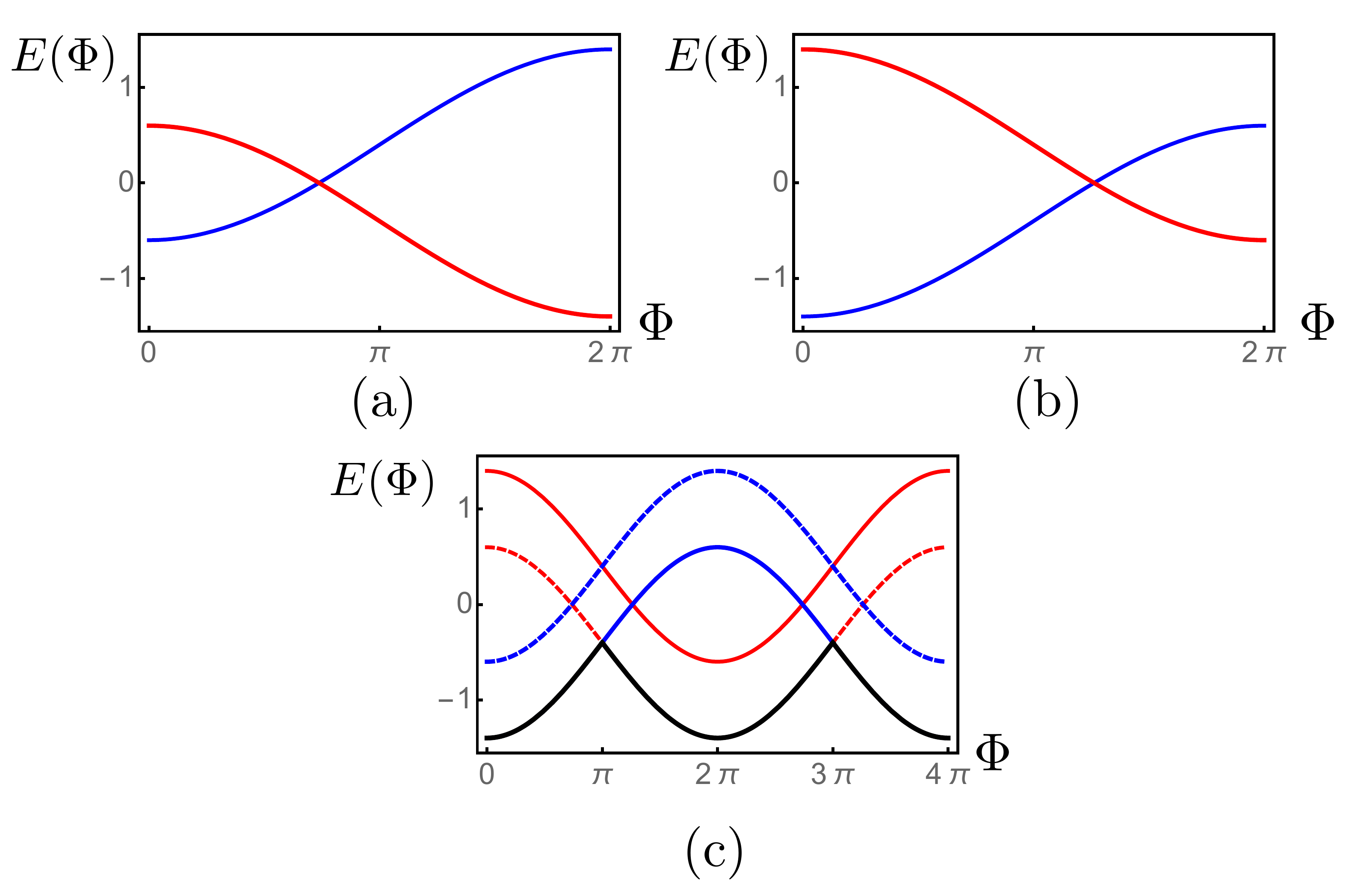}
	\caption{Schematic dependence of the many-body energy spectrum on magnetic flux (in units of SC flux quantum) for fixed overall fermion parity (a) $P=1$ and (b) $P=-1$, and (c) fluctuating fermion parity. Red(blue) lines correspond to even (odd) fermion parity at the junction. (c) The case when fermion parity is not conserved. One can notice that the ground state energy (solid black line) is $2\pi$-periodic.}
	\label{fig:andreevspec}
\end{figure}

We now discuss a new experiment for the detection of the fractional Josephson effect using the ground-state properties of the system. Let us consider a mesoscopic ring shown in Fig.~\ref{fig:setup}, in which the global fermion parity $\mathcal{\hat P}=\gamma_1\gamma_2\gamma_3\gamma_4$ is fixed either by an appreciable charging energy or by galvanically isolating the nanowire-superconductor system so that external electron tunneling (quasiparticle poisoning) is prohibited energetically. Next we assume that the distances $L_{12}$ and $L_{34}$ are much larger than the superconducting coherence length $\xi$ so that $\delta E_{12}=\delta E_{34}\rightarrow 0$. (Otherwise, the flux periodicity of the current in the ring is $4\pi$ due to the coherent single electron tunneling, similar to the persistent currents in non-superconducting mesoscopic rings~\cite{Buttiker1983}, regardless whether the system is in the topologically trivial or non-trivial phase.) Using the constraint on global fermion parity $\gamma_1\gamma_4=-P\gamma_2\gamma_3$, the effective two-level Hamiltonian can be written in terms of fermion parity at the junction $i\gamma_2\gamma_3=1-2n$ with $n=0,1$:
\begin{equation}
	H_{P=1}=-\Big(E_J\cos\frac{\Phi}{2}+\delta E_{14}\Big)(2n-1).
	\label{}
\end{equation}
Without $\delta E_{14}$, the ground-state energy of the system is $2\pi$-periodic with $\Phi$. However, when $\delta E_{14}$ is finite, the ground state energies at $\Phi=0$ and $\Phi=2\pi$ are clearly different, see Fig.~\ref{fig:andreevspec}, and, thus, the supercurrent in the ring is not $2\pi$-periodic anymore as discussed in details in Sec.\ref{sec:splitting}. In other words, we predict that provided the charging energy of the ring and the splitting energy $\delta E_{14}$ are sufficiently large, one should be able to detect $4\pi$-periodic Josephson effect in the {\it dc} measurements. We note that a crucial requirement of our proposal is that global fermion parity is fixed. In the opposite case (i.e. global fermion parity is not conserved), the ground state becomes $2\pi$-periodic, see Fig.~\ref{fig:andreevspec}c.

So far we have completely neglected the superconducting phase slips which are important in mesoscopic structures. In order to take into account quantum phase fluctuations properly, we will use the bosonisation technique which is well-suited for the problem at hand. We will show that quantum phase slips affect the level crossing between different fermion parity states (i.e. the splitting energy $E_{12}$ becomes power-law dependent with the system size) and, as such, are detrimental for the fractional {\it ac} Josephson effect. On the other hand, quantum phase slips are relatively innocuous in the context of the fractional {\it dc} Josephson effect, and our microscopic results using the bosonisation technique are in qualitative agreement with those obtained using the simple toy model above.

\section{Josephson effect in Luttinger liquids}\label{sec:meanfield}
In this section we first review Josephson effect in a spinful Luttinger liquid coupled to a conventional $s$-wave superconductor~\cite{MaslovPRB1996, AffleckPRB2000, SeidelPRB2005, NobuhikoPRB2007} and obtain the spectrum of Andreev states as a function of superconducting phase difference across the junction. Next, we discuss a spinless Luttinger liquid coupled to spinless p-wave superconductors~\cite{BarbarinoNJP}. The latter supports Majorana zero-energy modes which are ultimately responsible for the change of fundamental periodicity of the Josephson current. Finally, we will review Josephson effect in more exotic structures involving parafermions.

\subsection{Josephson effect in a Luttinger liquid coupled to an s-wave superconductor.}\label{sec:swave}

Let us consider Luttinger liquid coupled to two bulk s-wave superconductors with the superconducting phase difference $\Phi$. After integrating out the superconducting degrees of freedom, the effective Hamiltonian for the system becomes ($\hbar=c=1$)
\begin{align}\label{eq:model_swave_fermion}
H&=H_{\rm NW}+ \int dx \left( \Delta (x)\psi_{\uparrow}(x)\psi_{\downarrow}(x)+{\rm H.c.}\right)\\
&H_{\rm NW}=\!\!\!\int \!\di x\, \psi_{\sigma}^\dag(x)\!\!\left(-\frac{\partial_x^2}{2m^*}\!-\!\mu\!\right)_{\!\!\sigma\sigma'}\!\!\!\psi_{\sigma'}(x)\nonumber
\end{align}
where $m^*$ and $\mu$ are the effective mass and chemical potential. One can introduce magnetic field into the Hamiltonian\eqref{eq:model_swave_fermion} via minimal coupling $-i\partial_x\rightarrow -i\partial_x+eA$ and then gauge it away so that it only appears in the superconducting pairing potential. Thus, the induced superconducting pair potential $\Delta(x)$ is given by
\begin{align}\label{eq:pairing}
\Delta (x)=
\begin{cases}
	\Delta_0 & x<0\\
	0 & 0<x<l\\
	\Delta_0e^{-i\Phi} & x>l
\end{cases}.
\end{align}
We will assume $\Delta_0>0$ in the following.
In a ring geometry, see Fig.~\ref{fig:setup}, the phase difference $\Phi$ can be related to the magnetic flux through the loop $\Phi=2\pi f/f_0$ with $f$ being the flux piercing the ring and $f_0=\frac{hc}{2e}$ is the SC flux quantum.


We now perform standard bosonization procedure for spinful fermions using the convention~\cite{Giamarchi:2004}:
\begin{equation}
\label{eqn:spinful-bosonization}
\psi_{r,\sigma}=\frac{1}{\sqrt{2\pi a}}e^{-\frac{i}{\sqrt{2}}[(r \phi_{\rho}-\theta_{\rho})+\sigma(r \phi_{\sigma}\!-\!\theta_{\sigma})]}
\end{equation}
where $r=\pm$ and $\sigma=\pm$ for right/left-moving fermion with $\up/\dn$ spin, and $a$ the lattice cutoff. The effective Hamiltonian now reads
\begin{align}\label{eq:model_swave}
H&=\int dx \sum_{\mu=\rho, \sigma}\frac{v_{\mu}}{2\pi}[K_{\mu}(\partial_x\theta_{\mu})^2+K_{\mu}^{-1}(\partial_x\varphi_{\mu})^2]\nonumber\\
&-\int_{x<0} dx \frac{\Delta_0}{2\pi a}\cos(\sqrt{2}\theta_\rho)\cos\sqrt{2}\varphi_\sigma\nonumber\\
&-\int_{x>l} dx \frac{\Delta_0}{2\pi a}\cos(\sqrt{2}\theta_\rho-\Phi)\cos\sqrt{2}\varphi_\sigma
\end{align}
Assuming that $\Delta_0$ is large, the proximity-induced terms constrain the values of  $\theta_\rho$ and $\varphi_\sigma$ to the minima of the cosine potential in the corresponding bulk superconductors
\begin{align}\label{eq:BC1}
	\theta_{\rho}(x<0)&= 0 \mbox{ and } \theta_{\rho}(x>l)= \frac{\pi \hat{J}_{\rho}\!+\!\Phi}{\sqrt{2}}\\
\varphi_{\sigma}(x<0)&= 0 \mbox{ and } \varphi_{\sigma}(x>l)= \frac{\pi \hat{N}_{\sigma}}{\sqrt{2}},
\end{align}
where $\hat{J}_{\rho}=\sum_{r, s}r N_{r,s}$ and $\hat{N}_{\sigma}=\sum_{r, s} s N_{r,s}$ with $N_{r,s}$ being the zero momentum component of the density operator with spin $s$ and chirality $r$. Thus, the problem has been effectively reduced to solving for the modes of Luttinger liquid subject to the above boundary conditions. Note that the allowed integer values for the operators $\hat{J}_{\rho}$ and $\hat{N}_{\sigma}$ have to obey certain constraints (superselection rules): $(-1)^{\hat{J}_{\rho}}=(-1)^{\hat{N}_{\sigma}}$ which immediately follows from Eq. \eqref{eq:model_swave}.

We now expand the bosonic fields in terms of the normal modes satisfying the above boundary conditions:
\begin{align}
\varphi_{\rho}(x)&=\sqrt{2}{\varphi}^{(0)}_{\rho}+\sqrt{{K_{\rho}}{}}\sum_{n>0}\frac{1}{\sqrt{n}}\cos\frac{\pi n x}{l}(a_{n\rho}+a_{n\rho}^\dag)\nonumber\\	
\theta_{\rho}(x)&=\frac{\pi J_{\rho}+\Phi}{\sqrt{2}}\frac{x}{l}+\frac{i}{\sqrt{K_{\rho}}}\sum_{n>0}\frac{1}{\sqrt{n}}\sin\frac{\pi n x}{l}(a_{n\rho}^\dag-a_{n\rho}).\nonumber\\
\varphi_{\sigma}(x)&=\frac{\pi N_{\sigma}}{\sqrt{2}}\frac{x}{l} + i \sqrt{K_{\sigma}} \sum_{n>0}\frac{1}{\sqrt{n}}\sin\frac{\pi n x}{l}(a_{n\sigma}^\dag-a_{n\sigma})\nonumber\\	
\theta_{\sigma}(x)&=\sqrt{2}\theta^{(0)}_{\sigma}+\frac{1}{\sqrt{K_{\sigma}}}\sum_{n>0}\frac{1}{\sqrt{n}}\cos\frac{\pi n x}{l}(a_{n\sigma}+a_{n\sigma}^\dag).\nonumber
\end{align}
Here $a_n$ and $a_n^\dag$ are the annihilation and creation operators for particle-hole excitations, satisfying the canonical commutation relation $[a_{m\mu}, a_{n\nu}^\dag]=\delta_{mn}\delta_{\mu\nu}$. The operators ${\varphi}^{(0)}_{\rho}$ and $\theta^{(0)}_{\sigma}$ represent the zero modes of the corresponding fields satisfying the commutation relations $[\varphi^{(0)}_{\rho}, \hat{J}_{\rho}]=i$ and $[\theta^{(0)}_{\sigma}, \hat{N}_{\sigma}]=-i$. Using the above normal-mode expansion, one can find the energy of the system:
\begin{align}
\mathcal{E}(N_\sigma, J_\rho)&=\frac{\pi}{4}\frac{v_{\sigma}}{K_\sigma l} N_{\sigma}^2+\frac{\pi }{4}\frac{v_{\rho}K_\rho}{l} \left(J_{\rho}+\frac{\Phi}{\pi}\right)^2\nonumber\\
&+\sum_{\mu=\rho, \sigma}\sum_{k >0} \omega_{\mu}(k)\left(n_{\mu}(k)+\frac{1}{2}\right),
\end{align}
where $\omega_{\mu}(k)=v_{\mu}\frac{\pi k}{l}$ and $n_{\mu}(k)=\langle a^\dag_{\mu}a_{\mu}\rangle $. One can notice that the partition function for the system factorizes into the product of the zero modes $Z_0$ and finite-energy excitations $Z_n$: $Z=Z_0 Z_n$ with only $Z_0$ being dependant on the flux $\Phi$
\begin{align}
Z_0=\sum_{N_{\sigma}, J_{\rho} \in \mathbb{Z}}e^{-\beta \mathcal{E}(N_\sigma, J_\rho)}.
\end{align}
Clearly, the sector with odd $N_\sigma$ is gapped out which constraints the values of $J_\rho$ to be even. Thus, the flux-dependent ground-state energy of the system becomes
\begin{align}
E_g(\Phi)=\min_{m \in \mathbb{Z}} \frac{\pi K_\rho v_{\rho}}{l} \left(m +\frac{\Phi}{2\pi}\right)^2.
\end{align}
Finally, one can obtain the expression for Josephson current using
\begin{align}
I_J(\Phi)=2 \frac{\partial E_g(\Phi)}{\partial \Phi}
\end{align}
One can see that the ground-state energy and the Josephson current through the junction are $2\pi$-periodic which is consistent with the previous studies of the Josephson effect in s-wave superconductors\cite{MaslovPRB1996}.

\subsection{ Fractional Josephson effect in topological $p$-wave superconductors.}\label{sec:pwave}
Next, we consider a case of topological p-wave superconductors and study Josephson effect in the presence of Majorana zero modes. Realistically a ``spinless nanowire" can be engineered in spin-orbit-coupled spin-$1/2$ quantum wires subject to an external Zeeman field~\cite{1DwiresLutchyn, 1DwiresOreg}. The spinless nanowire is then proximity-coupled to a bulk s-wave superconductors at $x<0$ and $x>l$. The Hamiltonian for the ``spinless nanowire" reads:
\begin{align}
\!\!\!H_{\rm NW}\!&=\!\!\!\int \!\di x\, \psi_{\sigma}^\dag(x)\!\!\left(\!-\!\frac{\partial_x^2}{2m^*}\!-\!\mu\!+\!i\alpha \sigma_y \partial_x\!+\!V_z\sigma_z\!\right)_{\!\!\sigma\sigma'}\!\!\!\psi_{\sigma'}(x), \nonumber\\
H_{\rm P}&=\int dx \left[ \Delta(x) \psi_{\uparrow} \psi_{\downarrow}
+\text{h.c.}\right].
\label{eq:H0_prox}
\end{align}
where $\alpha$ is the strength of the spin-orbit Rashba interaction and $V_z$ is the Zeeman splitting, and superconducting pairing is defined in Eq.\eqref{eq:pairing}. When chemical potential $\mu < V_z$, the nanowire is effectively spinless. The electron tunneling between the NW and the SC leads to the
proximity effect described by the Hamiltonian $H_{\rm P}$. The superconducting pairing potential $\Delta_0$ is assumed to
be a static classical field and quantum fluctuations of the superconducting phase are neglected.

Assuming that Zeeman gap is large, one performs standard bosonization procedure to find the effective Hamiltonian to be equivalent to a spinless nanowire coupled to spinless p-wave superconductors:
\begin{align}
  H&=\int \di x \frac{v}{2\pi}[K(\partial_x\theta)^2+K^{-1}(\partial_x\varphi)^2]\\
  &-\frac{\Delta_P}{2\pi a}\left(\int_{x<0}\di x\,\cos 2\theta+\int_{x>l}\di x\,\cos (2\theta-\Phi)\right)\nonumber.
  \label{}
\end{align}
The superconducting phase difference across the junction is $\Phi$. We assume the superconducting pairing potential is large (i.e. $\Delta_P$ is relevant and flows to strong coupling under  the renormalization group flow) and gaps out the Luttinger liquid in the region $x<0$ and $x>l$. In this limit, the values of $\theta$ are constraint to the minima of the cosine potential imposing the following boundary conditions for the LL in the region $0<x<l$:
\begin{equation}
  \theta(0)=0 \mbox{ and }  \theta(l)=\frac{\Phi +2\pi \hat{J}}{2}.
  \label{}
\end{equation}
where $\hat J=N_R-N_L$ with $N_{r}$ being the zero momentum component of the density operator with chirality $r$. One can expand the bosonic fields in terms of normal modes satisfying the boundary conditions:
\begin{equation}
  \begin{gathered}
  \varphi(x)={\varphi}^{(0)}+\sqrt{{K}{}}\sum_{n>0}\frac{1}{\sqrt{n}}\cos\frac{\pi n x}{l}(a_n+a_n^\dag)\\	
	\theta(x)=\frac{\Phi +2\pi \hat{J}}{2}\frac{x}{l}+\frac{i}{\sqrt{K}}\sum_{n>0}\frac{1}{\sqrt{n}}\sin\frac{\pi n x}{l}(a_n^\dag-a_n).
  \end{gathered}
  \label{eq:normal_mode_spinless}
\end{equation}
Here $a_n$ and $a_n^\dag$ are annihilation and creation operators for particle-hole excitations, satisfying the canonical commutation relation $[a_m, a_n^\dag]=\delta_{mn}$; ${\varphi}^{(0)}$ is the zero mode of the $\varphi$ field and is conjugate to $\hat J$: $[{\varphi}^{(0)}, \hat{J}]=i$. After substituting \eqref{eq:normal_mode_spinless} into the effective Hamiltonian for the junction, one finds
\begin{equation}
	\begin{gathered}
    H=\frac{\pi v K}{2 l}\left(\hat{J}+\frac{\Phi}{2\pi}\right)^2+\sum_{n>0} \frac{v\pi n}{l}\left(a_n^\dag a_n+\frac{1}{2}\right),\\
	\end{gathered}
  \label{}
\end{equation}
One can now easily find the flux dependent part of the ground state energy
\begin{align}
E_g(\Phi)=\min_{J \in \mathbb{Z}}\frac{\pi vK}{2l}\left( J+\frac{\Phi}{2\pi} \right)^2
\end{align}
We remind that different parity of $J=N_R-N_L$ actually corresponds to different parity of electron number operator $N=N_R+N_L$. If electron number in the junction is conserved, $J$ should be restricted to either even or odd sectors, i.e. $J=2m+\frac{1-(-1)^N}{2}$. Thus, the ground state energy as well as the current are $4\pi$-periodic: $E_g(\Phi)=E_g(\Phi+4\pi)$. On the other hand, if there are processes allowing to change the fermion parity in the junction, the ground-state energy is $2\pi$-periodic. In practice, one should define the time scale associated with such processes $\tau_P$. When $t\gg \tau_P$ (dc limit), the Josephson current in spinless superconductors is $2\pi$-periodic and, in this sense, is similar to the Josephson effect in conventional s-wave superconductors. However, if measured at $t \ll  \tau_P$ (ac limit), the fundamental periodicity of the Josephson current is $4\pi$, and, thus, one could distinguish between the topological (spinless p-wave) and non-topological (spinful s-wave) junctions. This is why this phenomenon in the literature was ``coined" as fractional ac Josephson effect~\cite{1DwiresKitaev, Kwon}.

 \begin{figure}[!t]
	 \centering
	 \includegraphics[width=0.8\columnwidth]{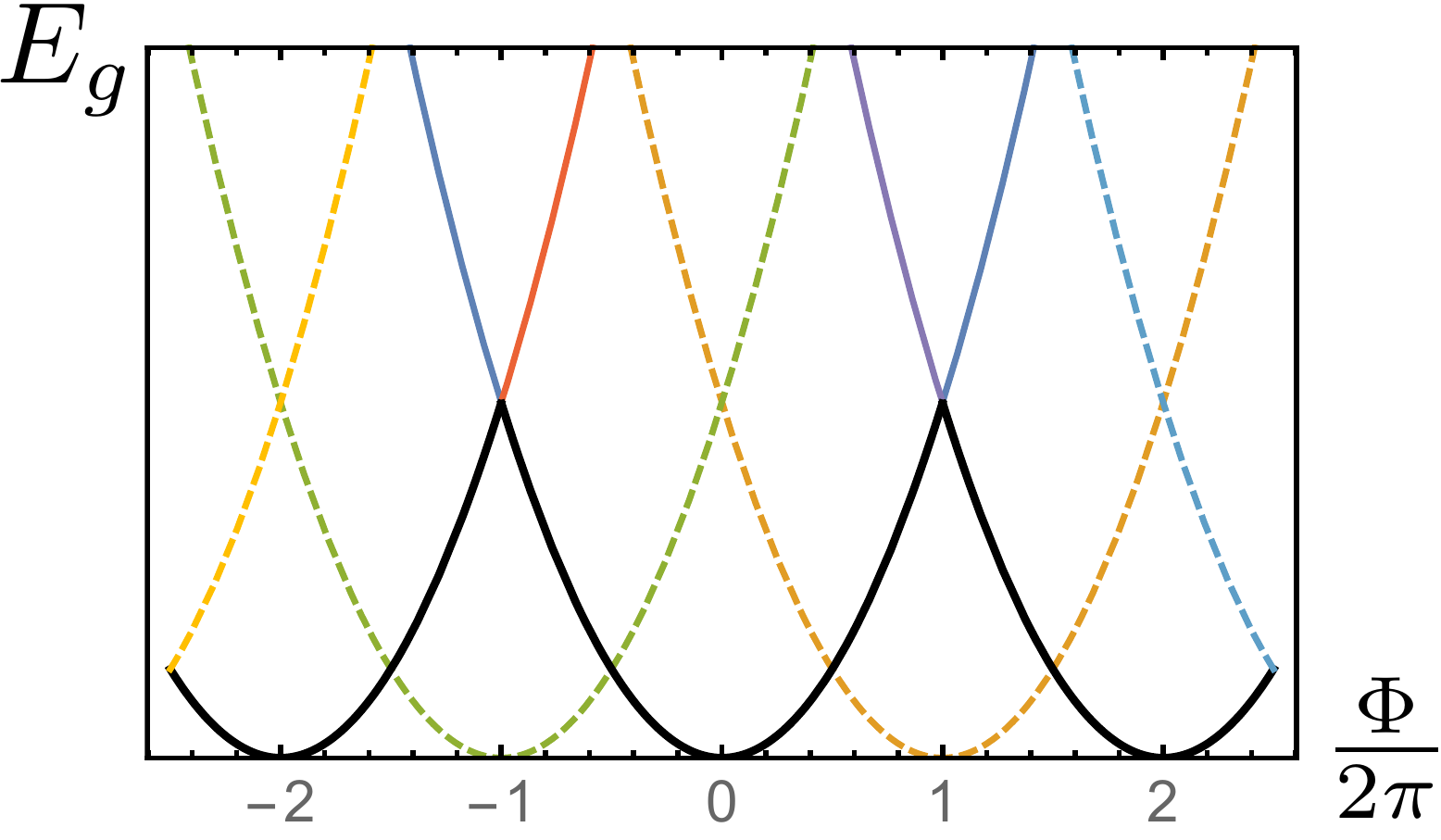}
	 \caption{Ground state energy of a spinless Josephson junction (black thick line) as a function of the phase difference. Each parabolic curve corresponds to distinct values of $J$(solid lines correspond to even $J$ and dashed ones are odd). }
	 \label{fig:energy}
 \end{figure}

\subsection{ Fractional Josephson effect in parafermion systems.}
\label{sec:parafermion}
We now review fractional Josephson effect in the presence of parafermionic zero modes. Instead of a spinless Luttinger liquid discussed above in the Majorana context, we now consider gapless edge modes of a fractional topological insulator, or the counter-propagating edge modes along a trench in a ${\nu}=\frac{1}{m}$ Laughlin state. The fundamental excitations in this case are fractionalized quasiparticles characterized by the fields $\varphi$ and $\theta$ satisfying the following commutation relations:
\begin{equation}
	[\varphi(x),\theta(x')]=\frac{i\pi}{m}\Theta(x-x').
  \label{eq:comm_paraf}
\end{equation}
with $m >1$. The effective Hamiltonian for the edge reads
\begin{equation}
	H=\frac{v m}{2\pi}\int \di x [K(\partial_x\theta)^2+K^{-1}(\partial_x\varphi)^2].
	\label{eqn:luttinger}
\end{equation}
The physical electron operators in this effective theory are given by $\psi_{R/L}=\frac{1}{\sqrt{2\pi a}}e^{-im(\pm\varphi-\theta)}$, and, thus, the proximity-induced superconductivity due to spinless p-wave superconductor for $x<0$ and $x>l$ can be taken into account using the following effective Hamiltonian:
\begin{align}
H_{\Delta}=&-\frac{\Delta_P}{2\pi a}\left(\int_{x<0}\di x\,\cos 2m\theta+\int_{x>l}\di x\,\cos (2m\theta-\Phi)\right)
\end{align}
Assuming that $\Delta_P$ is large, we once again consider Luttinger liquid Hamiltonian confined in the region $0<x<l$ subject to the boundary conditions:
\begin{align}
	 \theta(0)=0 \mbox{ and } \theta(l)=\frac{\Phi+2\pi \hat{J}}{2m}.
	 \label{}
\end{align}
Following similar analysis as in the previous section,  we find the normal mode expansion for the fields $\theta$ and $\varphi$ is given by
\begin{align}
\varphi(x)&={\varphi}^{(0)}+\sqrt{\frac{K}{m}}\sum_{n>0}\frac{1}{\sqrt{n}}\cos\frac{\pi n x}{l}(a_n+a_n^\dag)\nonumber\\	
\theta(x)&=\frac{\Phi +2\pi \hat{J}}{2m}\frac{x}{l}+\frac{i}{\sqrt{K m}}\sum_{n>0}\frac{1}{\sqrt{n}}\sin\frac{\pi n x}{l}(a_n^\dag-a_n)\nonumber.
\end{align}
The fields $\theta(x)$ and $\varphi(x)$ should satisfy the commutation relations \eqref{eq:comm_paraf} and have fundamental periodicity of $2\pi$. The latter imposes a constraint on the values of $J$ requiring that $J=2m k+n_J$. After substituting above expressions for $\theta(x)$ and $\varphi(x)$ into Eq.\eqref{eqn:luttinger}, the ground-state energy for the system is given by
\begin{equation}
	 \begin{split}
	 E_g(\Phi)&=\min_{k}\frac{2\pi  v K}{ml}\left(k+\frac{n_J}{2m}+\frac{\Phi}{4\pi m} \right)^2\\
	 \end{split}
	 \label{}
 \end{equation}
where $n_J \in \mathbb{Z}_{2m}$ labels different topological sectors. One can see that ground state energy is $4\pi m$ periodic $E_g(\Phi)=E_g(\Phi+4\pi m)$.

\section{Fractional ac Josephson effect in Number-Conserving Systems}\label{sec:number_cons}

So far we have included superconductivity at the mean field level neglecting quantum phase fluctuations, which is appropriate when a nanowire is coupled to bulk superconductors. In other words, this approximation implies that the particle number is not conserved corresponding to the grand canonical ensemble. In mesoscopic structures, however, particle number fluctuations might be suppressed by the charging energy, and, thus, it is interesting to investigate Josephson effect in particle number conserving systems. Given that particle number and superconducting phase are conjugate variables, one needs to take into account strong quantum fluctuations of the superconducting phase in particle number conserving systems (canonical ensemble). This fact is particularly important in one-dimensional systems, where $\mathrm{U}(1)$ symmetry can not be spontaneously broken due to the Mermin-Wagner theorem. Therefore, in the following we consider the fractional Josephson effect in a model where a nanowire is coupled to the quasi-long-range ordered superconductor (QSC) with strongly fluctuating SC phase.

We now consider an attractive Hubbard model, where the spin backscattering caused by the electron interaction is marginally relevant and flows to strong coupling, resulting in the formation of the Luther-Emery phase~\cite{LutherPRL1974} with a finite spin gap but no charge gap, and therefore can be thought as a one-dimensional analogy of an s-wave superconductor. After the bosonization, the Hamiltonian for the superconducting wire reads
\begin{align}
H_{\rm SC}&=H^{(\rho)}_{\rm SC}+H^{(\sigma)}_{\rm SC}\label{eq:H_AH_total} \\
H^{(\rho)}_{\rm SC}&=\frac{v_\rho}{2\pi}\int \di x \left[K_{\rho} (\partial_x \theta_{\rho})^2+ K_{\rho}^{-1}(\partial_x \phi_{\rho})^2\right] \label{eq:H_AH_rho} \\
H^{(\sigma)}_{\rm SC}&=\frac{v}{2\pi}\int \di x \left[K_{\sigma} (\partial_x \theta_{\sigma})^2+ K_{\sigma}^{-1}(\partial_x \phi_{\sigma})^2\right] \label{eq:H_AH_sigma} \\
&-\frac{2|U|}{(2\pi a)^2}\int \di x \cos(2\sqrt{2}\varphi_{\sigma}) \nonumber
\end{align}
where $v_F$, $a$ and $U$ are the Fermi velocity, the effective cutoff length and the interparticle interaction potential, respectively. The physics of this quasi-long-range superconducting wire can be understood in terms of fluctuating Cooper pairs having algebraically-decaying correlations. The cosine term in Eq.~\eqref{eq:H_AH_sigma} is marginally relevant so that the spin field $\varphi_\sigma$ is pinned to the classical minima, opening a spin gap $\Delta_\text{s}\propto e^{-\frac{\pi v_\sigma}{|U|}}$ in the QSC. As shown below, spin gap is crucial for our construction since it prohibits single-electron tunneling to the QSC at low energies. Thus, Cooper pair tunneling is the dominant tunneling process between the nanowire and the superconductor. One can relate the parameters of the above model for the Luther-Emery phase to that of a quasi-1D superconductor: $K_\rho=2\pi \sqrt{A_w \rho_s \kappa}$ and $v_\rho=\sqrt{A_w \rho_s /\kappa}$ with $A_w$, $\rho_s$ and $\kappa$ being the cross-sectional area of the superconductor, the superfluid stiffness and the compressibility, respectively.

Our theoretical model also involves the Hamiltonian for the nanowire $H_\text{NW}$~\eqref{eq:H0_prox}, in which we once again assume that the Zeeman field is large so that the nanowire is in the ``spinless regime" with only the lowest band occupied. The nanowire and the QSC are coupled by the single particle tunneling term:
\begin{equation}
	H_T=t\sum_\sigma\int\di x\, (\psi_\sigma^\dag\eta_\sigma+\text{h.c.}).
	\label{}
\end{equation}
Here $\psi_\sigma$ and $\eta_\sigma$ are electron annihilation operators in the nanowire and QSC, respectively. After the bosonization, we arrive at following Hamiltonian:
\begin{align}
H=H_{\rm NW}(\theta, \varphi)+H^{(\rho)}_{\rm SC}(\theta_\rho, \varphi_\rho)+H^{(\sigma)}_{\rm SC}(\theta_\sigma, \varphi_\sigma)+H_T.
\end{align}
Given that single-electron tunneling into the superconducting wire is suppressed due to the presence of the spin gap $\Delta_{s}$, the dominant contribution to the low-energy effective action comes from pair tunneling. The perturbative expansion in $t$ to second order leads to the following imaginary-time action
\begin{align}\label{eq:pair_hopping1}
S_{\rm PH}\!&=\!- t^2\sum_{\sigma}\!\!\int\!\! dx d \tau dx' d \tau'\! \!\\
&\left[ \psi^\dag_{\sigma}(x,\tau)\psi^\dag_{-\sigma}(x',\tau')\eta_{\sigma}(x,\tau)\eta_{-\sigma}(x',\tau') \!+\! \text{h.c.} \right].\nonumber
\end{align}
Given that the spin field $\varphi_{\sigma}$ orders as a result of the last term in Eq. \ref{eq:H_AH_sigma}, the dual field $\theta_{\sigma}$ is strongly disordered, and its correlation function decays exponentially:
\begin{equation}
	\langle e^{-\frac{i}{\sqrt 2} \theta_{\sigma}(x,\tau)} e^{\frac{i}{\sqrt 2} \theta_{\sigma}(0,0)} \rangle \sim \frac{ae^{-\frac{\Delta_\text{s}}{v_\sigma}\sqrt{x^2+v_\sigma^2\tau^2}}}{\sqrt{x^2+v_\sigma^2 \tau^2}} .
	\label{}
\end{equation}
This allows us to simplify the action \eqref{eq:pair_hopping1} and make a local approximation
\begin{align}\label{eq:pair_hopping}
S_{\rm PT}\!&\approx \!- \frac{\Delta_P}{(2\pi a)^2} \int \di\tau \int\di x\, \cos\left(2\theta-\sqrt{2} \theta_{\rho}\right),
\end{align}
which is valid in the long-time limit $|\tau-\tau'| \gg \Delta_\text{s}^{-1}$. Here the Cooper
pair tunneling amplitude $\Delta_P$ is given by $\Delta_P \sim \frac{t^2}{\Delta_\text{s}} \frac{\alpha p_F}{\sqrt{(\alpha p_F)^2+V_z^2}}$. The derivation above can be straightforwardly generalized to the case of fractionalized Luttinger liquid discussed in Sec. \ref{sec:parafermion} where the index $m$ corresponds to a specific edge theory, i.e. $m=1$ represents Majorana case whereas $m > 1$ corresponds to a specific parafermion model. After some algebra, one finds that the effective action for the tunneling between QSC and LL now reads
\begin{align}
S_{\rm PT}\!&\approx \!- \frac{\Delta_P}{(2\pi a)^2} \int \di\tau \int\di x\, \cos\left(2m\theta-\sqrt{2} \theta_{\rho}\right).
\end{align}

We also notice that for $m>1$, it is not physical to consider a finite fractionalized liquid since it exists on the edge of a 2D system which has no boundaries. We therefore have to induce a distinct gap on the edge to terminate the paired topological regions e.g by a backscattering term $\psi_R^\dag\psi_L+\text{h.c}$, which becomes $\cos 2m\varphi$ after bosonization. We assume this is the case for $m>1$ in the following discussion.

Finally, one arrives at the effective Hamiltonian for the system of interest:
\begin{align} \label{}
	{H}&={H}_\text{NW}+{H}_\text{SC}^\rho+{H}_\text{PT}\\
	{H}_\text{NW}&=\frac{v}{2\pi}\int \di x [K(\partial_x\theta- A)^2+K^{-1}(\partial_x\varphi)^2]\nonumber\\
	{H}_\text{SC}^\rho&=\frac{v_\rho}{2\pi}\int \di x [K_\rho(\partial_x\theta_\rho-\sqrt{2} A)^2+K_\rho^{-1}(\partial_x\varphi_\rho)^2]\nonumber\\
{H}_\text{PT}&=-\frac{\Delta_P}{2\pi a}\int \di x \cos(2m\theta - \sqrt{2}\theta_\rho).\nonumber
\end{align}
Here we also introduced the vector potential $A$ due to the out-of-plane magnetic field, see Fig.\ref{fig:setup}. Henceforth, we assume that Cooper-pair tunneling term is large so that it ``locks'' the phase difference between the edge modes/nanowire and the QSC $2m\theta - \sqrt{2}\theta_\rho$.
 \footnote{The scaling dimension of the pair tunneling term is $\frac{1}{2}K_\rho^{-1}+m^2K^{-1}<2$, and it is relevant in the Majorana case ($m=1$)  and flows to strong coupling limit.}.

Following Ref.~\onlinecite{Fidkowski2011}, we now review the topological degeneracy in the wires/edge proximity-coupled to QSC. Given that the total number of electrons is conserved, the minimal setup with topologically protected ground-state degeneracy involves four domain walls (i.e. two separated nanowires) coupled to the same QSC, see Fig.\ref{fig:setup}. In the topological regions, $\Theta=\theta-\frac{\theta_\rho}{\sqrt{2}m}$ are pinned to the classical minima and we denote its value in the first and second wire by $\Theta_1$ and $\Theta_2$, respectively.
Naively, the pair tunneling leads to superficially $(2m)^2$-fold degenerate ground state manifold: $(\Theta_1, \Theta_2)=\frac{\pi}{m}(n_1,n_2), n_{1,2}\in \{0,1,\dots,2m-1\}$. However, one needs to be more careful and study the moduli space of the phase variables. Indeed, since the two wires are coupled to the same QSC, we are allowed to make a global gauge transformation $\theta_\rho\rightarrow\theta_\rho+2\pi$, which leaves the ground state invariant. This shows that we need to make the following identification: $(n_1, n_2)\sim (n_1+{k},n_2+k), k\in\mathbb{Z}$. Therefore, we can fix $\Theta_1=0$, and different ground states are labeled by $\Theta_2$ and we have $2m$-fold ground state degeneracy.

We now study Josephson current for the setup shown in Fig. \ref{fig:setup}. We consider a ring of length $L$ with $0\leq x<L$, and assume that the nanowire covers $[0,l]\cup [L-l-l', L]$, and the QSC covers $[0, L-l']$ with $l'$ being the length of the junction. Using this coordinate system, the four Majorana zero modes $\gamma_1,\gamma_2,\gamma_3,\gamma_4$ in Fig. \ref{fig:setup} are located at $x=L-l-l', L-l', 0, l$ respectively.
The vector potential can be chosen to be $A=\frac{f}{L}$ where $f$ is the magnetic flux threading the loop.  We first perform the following gauge transformation to eliminate the vector potential from the Hamiltonian:
\begin{equation}
	\begin{gathered}
	\eta_\sigma(x)\rightarrow \eta_\sigma(x) e^{-\frac{i\Phi x}{2L}},\\
  \psi(x)\rightarrow
  \begin{cases}
	\psi_\sigma(x)e^{-\frac{i\Phi x}{2L}} & L-l-l'\leq x < L\\
	\psi_\sigma(x)e^{-\frac{i\Phi(x+L)}{2L}} & 0\leq x\leq l
  \end{cases}	
	\end{gathered}
  \label{}
\end{equation}
where $\Phi=2\pi f/f_0$ with $f$ being the magnetic flux piercing the ring and $f_0=\frac{hc}{2e}$. It can be readily checked that this transformation is continuous at $x=0$ and, thus, fermion operators are single-valued everywhere (i.e. this transformation does not introduce a branch cut in the fermionic fields).

However, the pair tunneling terms are affected in a nontrivial way and change under gauge transformation as
\begin{equation}
\eta_{\uparrow}^\dag \eta_{\downarrow}^\dag \psi_{}\psi_{}\rightarrow
  \begin{cases}
	\eta_{\uparrow}^\dag \eta_{\downarrow}^\dag \psi_{}\psi_{} & L-l-l'\leq x\leq L-l'\\
	e^{i\Phi}\eta_{\uparrow}^\dag \eta_{\downarrow}^\dag \psi_{}\psi_{} & 0\leq x\leq l.
  \end{cases}
  \label{}
\end{equation}
Upon standard bosonization, the modified pair tunneling terms are given by
\begin{multline}
  H_\text{PT}=-\frac{\Delta_P}{2\pi a} \int_0^{l}\di x\,\cos (2m\theta-\sqrt{2}\theta_\rho-\Phi)\\-\frac{\Delta_P}{2\pi a} \int_{L-l-l'}^{L-l'}\di x\,\cos (2m\theta-\sqrt{2}\theta_\rho)
  \label{}
\end{multline}
Assuming that $\Delta_P$ term is large, one can approximate the phase difference $\theta-\frac{\theta_\rho}{\sqrt{2}m}$ by its values at the minima of the cosine potential:
\begin{equation}
\theta-\frac{\theta_\rho}{\sqrt{2}m}=
	\begin{cases}
	0 & L-l-l'\leq x\leq L-l'\\
	\chi & 0\leq x\leq l,
	\end{cases}
	\label{eq:mintheta}
\end{equation}
where $\chi=\frac{\Phi+2\pi \hat{J}}{2m}$ and $J \in \mathbb{Z}$. As discussed in Sec. \ref{sec:parafermion}, $J \text{ mod }2m$ is a conserved quantity, corresponding to different superselection sectors.

We now calculate the ground state energy and its dependence on the magnetic flux $f$. It is convenient to integrate out $\varphi$ and $\varphi_\rho$ fields and write the partition function in terms of the imaginary-time effective action.
\begin{equation}
	\begin{split}
		{S}=&\int_0^\beta\di \tau\,\Big[\int_{L-l-l'}^{L+l} \di x\,\frac{mK}{2\pi v}[(\partial_\tau\theta)^2+v^2(\partial_x\theta)^2]\\
		&+\int_{0}^{L-l'} \di x\,\frac{K_\rho}{2\pi v_\rho}[(\partial_\tau\theta_\rho)^2+v_\rho^2(\partial_x\theta_\rho)^2]\\
	&-\frac{\Delta_P}{2\pi a} \int_0^{l}\di x\,\cos (2m\theta-\sqrt{2}\theta_\rho-\Phi)\\
&-\frac{\Delta_P}{2\pi a} \int_{L-l-l'}^{L-l'}\di x\,\cos (2m\theta-\sqrt{2}\theta_\rho)\Big]
	\end{split}
	\label{}
\end{equation}
 It is easy to see that the system is gapless in particle-number conserving setup contrary to the previous case discussed in Sec.\ref{sec:meanfield}. Indeed, the combination of the fields $2m\theta+\sqrt 2 \theta _\rho$ is free to fluctuate. Given that the combination of the fields $2m\theta-\sqrt 2 \theta _\rho$ is pinned in the topological regions, one can integrate them out. As a result, we can impose the following constraint:
\begin{equation}
  \frac{1}{\sqrt{2}m}\partial_{\tau/x}\theta_\rho=\partial_{\tau/x}\theta.
\end{equation}
Thus, the problem reduces to that of an inhomogeneous single-component Luttinger liquid:
 \begin{equation}
	S=\int_0^L\di x \di \tau \frac{\tilde{K}(x)}{2\pi \tilde{v}(x)}\left[ (\partial_\tau\tilde{\theta})^2+\tilde{v}^2(x)(\partial_x\tilde{\theta})^2 \right]
  \label{eqn:lut2}
\end{equation}
where the phase field $\tilde{\theta}$ is defined as
\begin{equation}
  \tilde{\theta}=
  \begin{cases}
	  \frac{1}{\sqrt{2}m}\theta_\rho & 0<x<L-l'\\
	  \theta & L-l'\leq x\leq L
  \end{cases}
  \label{}
\end{equation}
The Luttinger parameters $\tilde{K}(x)$ and velocity $\tilde{v}(x)$ are given by:
\begin{equation}
  \tilde{v}(x)=
  \begin{cases}
	v_+ & 0<x<l\text{ or } L-l-l'<x<L-l'\\
	v & L-l-l'<x<L\\
	v_\rho & l<x<L-l-l'
  \end{cases},
  \label{}
\end{equation}
and
\begin{equation}
  \tilde{K}(x)=
  \begin{cases}
	K_+ &  0<x<l\text{ or } L-l-l'<x<L-l'\\
	mK & L-l-l'<x<L\\
	2m^2K_\rho & l<x<L-l-l'
  \end{cases}
  \label{}
\end{equation}
with $v_+,K_+$ being
\begin{align}
	K_+&=\sqrt{K_\rho^2+\frac{K^2}{4m^2}+\frac{KK_\rho}{2m}\left( \frac{v}{v_\rho}+\frac{v_\rho}{v} \right)},\\
	v_+&=\sqrt{vv_\rho\frac{vK+2mv_\rho K_\rho}{v_\rho K+2mvK_\rho}}.
\end{align}

In order to calculate partition function, one needs to specify boundary conditions for the field $\tilde\theta$ at $x=0$ and $x=L-l'$. Since QSC terminates at these points, the appropriate boundary conditions are
\begin{equation}
	\partial_x\theta_\rho(x)=0 \text{ for } x=0, L-l'.
  \label{}
\end{equation}
The field $\theta$ should be continuous at $x=0, L-l'$ so that there are no singularities in the effective action.
\begin{equation}
	\theta(x^-)=\theta(x^+) \text{ for } x=0, L-l'.
  \label{}
\end{equation}
In terms of the field $\tilde{\theta}$, these boundary conditions translate to
\begin{equation}
	\begin{gathered}
		\tilde{\theta}(L-l+0^+,\tau)=\tilde{\theta}(L-l-0^+,\tau)\\
	\tilde{\theta}(0^+,\tau)=\tilde{\theta}(0^-,\tau)-\chi,
	\end{gathered}
	\label{}
\end{equation}
where we used Eq.\eqref{eq:BC1}. Thus, the modified field $\tilde{\theta}$ satisfies twisted boundary conditions at $x=0$. We should also include winding numbers of $\tilde{\theta}$ in the imaginary-time direction:
\begin{equation}
	\tilde{\theta}(x,\beta)=\tilde{\theta}(x,0)+2\pi mM.
	\label{}
\end{equation}
However, one needs to be careful about the values of $M$. Because of the pair tunneling term, $M$ must be an integer to preserve the periodicity of $\theta_\rho$. It is easy to see that to minimize the action, we can write
\begin{equation}
	\tilde{\theta}(x,\tau)=\frac{2\pi mM}{\beta}\tau+\tilde{\theta}(x),
	\label{}
\end{equation}
and the action evaluates to
\begin{equation}
	S=\frac{2\pi m^2}{\beta}\sum_i\frac{K_il_i}{v_i}M^2+\beta\int_0^L\di x\,\frac{\tilde{K}(x)\tilde{v}(x)}{2\pi}(\partial_x\tilde{\theta})^2.
	\label{eqn:action2}
\end{equation}
Now one can minimize the action~\eqref{eqn:action2} for a fixed $\chi$. Since $K(x), v(x)$ are piecewise constant, the action is minimized by piecewise-linear $\tilde{\theta}$.  To calculate partition function, one needs to find the stationary solution (i.e. $\partial_\tau\tilde{\theta}=0$) satisfying the following constraints. To simplify the notations, we define $x_0=0, x_1=l, x_2=L-l-l', x_3=L-l', x_4=L$ as the locations of the interfaces where $v$ and $K$ are discontinuous and the field $\theta(x_i)$ are given by
\begin{equation}
  \begin{gathered}
	  \tilde{\theta}(x_0)=0\equiv\theta_0\\
	  \tilde{\theta}(x_i)=\theta_i, 1\leq i\leq 3\\
	  \tilde{\theta}(x_4)=\chi\equiv\theta_N.
\end{gathered}
  \label{}
\end{equation}
Since we restrict ourselves to the space of piecewise linear functions, $\tilde{\theta}$ is then uniquely specified once all the values at $\{x_i\}_{i=0}^3$ are determined. Thus, the action we want to minimize is given by
\begin{equation}
  S[\{\theta_i\}]=\frac{\beta}{2\pi}\sum_{i=1}^N\frac{v_i K_i}{l_i}(\theta_i-\theta_{i-1})^2.
  \label{}
\end{equation}
where $l_i=x_i-x_{i-1}$ and $\beta$ is the inverse temperature.

After minimizing above expression with respect to $\tilde{\theta}(x_i)=\theta_i, 1\leq i\leq 3$, the minimum of the action is given by
\begin{equation}
	S_\text{min}=\frac{2\pi m^2 C_\text{loop}}{\beta}M^2+\frac{2\pi\beta}{m^2 L_{\rm loop}}\left( J+\frac{\Phi}{2\pi} \right)^2.
  \label{}
\end{equation}
where $J=2mk+n_J$ with $k \in \mathbb{Z}$ and $n_J$ being a number corresponding to different topological sectors, i.e. for $m=1$ this number simply denotes fermion parity. The total inductance of the loop $L_{\rm loop}=\sum_{i=1}^4 \frac{l_i}{ v_i K_i}$ is simply given by the sum of all individual inductances (inductors in series rule), and the total capacitance $C_\text{loop}=\sum_{i=1}^4 \frac{K_i l_i}{v_i}$. The partition function is obtained by summing over $k$:
\begin{equation}
	\begin{split}
	\mathcal{Z}=\sum_{k,M}\exp&\Big[-\frac{2\pi m^2 C_\text{loop}}{\beta}M^2\\
	&-\frac{2\pi\beta}{m^2L_{\rm loop}}\left(k+\frac{n_J}{2m}+\frac{\Phi}{4\pi m}\right)^2\Big].
	\end{split}
	\label{}
\end{equation}
Now we use the Poisson summation formula to rewrite the sum over $M$:
\begin{equation}
	\sum_M\exp\Big(-\frac{2\pi m^2 C_\text{loop}}{\beta}M^2\Big)=\sum_N\exp\Big(-\frac{\pi \beta}{2m^2C_\text{loop}}N^2\Big).
	\label{}
\end{equation}
Therefore, we obtain the ground-state energy for the system in a fixed $N, J$ sector (here we restored the units):
\begin{equation}
	E_\mathrm{g}^{N,J}(\Phi)=\frac{\pi e^2}{2m^2C_\text{loop}}N^2+\min_{J \in \mathbb{Z}}\frac{2 f_0^2}{\pi m^2 L_{\rm loop}}\left(J+\frac{\Phi}{2\pi}\right)^2.
	\label{eq:Jos_fixed}
\end{equation}
Eq.~\eqref{eq:Jos_fixed} is one of the main results of this paper showing that for $m=1$ the ground state energy is still $4\pi$ periodic provided fermion parity is conserved. We can now recover the previous results obtained for a bulk superconductor. Indeed, with the increase of the number of transverse channels $N_{\rm ch}$ in the QSC, quantum fluctuations are suppressed. In the limit $K_\rho\propto N_{ch} \rightarrow \infty$, the capacitance and inductance for the loop become $C_\text{loop}\rightarrow \infty$ and $L_{\rm loop}\approx \frac{l'}{v m K }$, and, thus, we recover the results of Sec. \ref{sec:parafermion}. Since $J$ and $N$ commute, the first term is just an additive constant corresponding to different particle number in the ring which is assumed to be fixed henceforth.

\section{Topological Degeneracy Splitting}\label{sec:splitting}

\begin{figure}[t]
	\centering
	\includegraphics[width=\columnwidth]{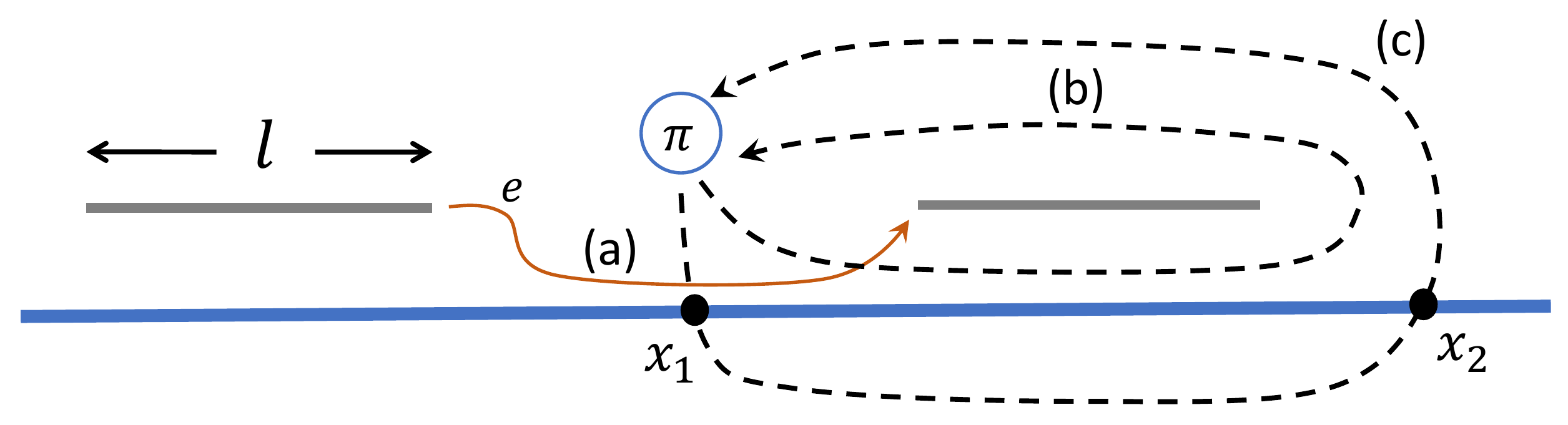}
	\caption{Illustration of degeneracy splitting processes. (a) electron tunneling between two nanowires through the QSC. (b) quasiparticle tunneling across the topological region. (c) Electron backscattering in the QSC.}
	\label{fig:splitting}
\end{figure}

An important aspect of a topological phase is its ground-state degeneracy. In a finite-size system, the degeneracy is lifted by certain instanton events which connect different topological sectors. In the case of a bulk superconductor, such events can be understood in term of the overlap of the MZM wave functions. In the interacting system considered here the splitting calculation becomes more subtle. One can identify the processes which are present in the BdG formalism and the ones that are not, i.e. generated by quantum fluctuations. Possible sources of the topological degeneracy splitting can be classified into three types, see Fig.~\ref{fig:splitting}: (a) electron tunneling between the nanowires through the QSC, (b) quasiparticle tunneling across the topological region, (c) splitting caused by electron backscattering in the QSC.

The splitting process (a) is due to the virtual tunneling of a single electron through the QSC between the topological wires (depicted in Fig. \ref{fig:splitting} as the process (a)). We obtain this term in the perturbative expansion of the action in the single-fermion tunneling amplitude $t$:
\begin{align}
	  H_{S2}&=-\frac{2t^2}{\Delta_\text{s}}
	{e^{-\Delta_\text{s}|x-x'|/v_F}}\\
  &\times\cos\Big[m\theta(x')-\frac{\theta_\rho(x')}{\sqrt{2}}+m\theta(x)-\frac{\theta_\rho(x)}{\sqrt{2}}-\frac{\Phi}{2}\Big]\nonumber\\
  &\times\cos \Big[m\varphi(x)+\frac{\varphi_\rho(x)}{\sqrt{2}}-m\varphi(x')-\frac{\varphi_\rho(x')}{\sqrt{2}}\Big]\nonumber
  \label{}
\end{align}
where $x, x'$ are the ends of the nanowire at $x=l$ and $x'=L-l-l'$, see Fig.\ref{fig:setup}. Using Eq. \eqref{eq:mintheta}, one finds that $H_{S2}$ is given by $H_{S2}=\delta E_a\cos(\pi J)$ with
\begin{align}
\!\delta E_a\!=-\frac{2t^2}{\Delta_\text{s}}
	\exp \left(-\frac{\Delta_\text{s}|x\!-\!x'|}{v_F}\!-\!\frac{1}{4K_\rho}\!\log\left|\frac{x\!-\!x'}{a}\right|\right).
\end{align}
Here we used the fact that $\varphi(x)$ and $\varphi(x')$ are pinned by the boundaries. For nanowires this is imposed by the boundary conditions at the ends of the nanowire whereas for parafermion setup the induced backscattering terms $\cos 2m\varphi$ effectively pin fields $\varphi(x)$ and $\varphi(x')$, see Refs. \onlinecite{Clarke12, Lindner12, cheng2012} for more details.

The splitting of the ground state degeneracy due to the tunneling of a fundamental topological charge can be obtained by the instanton calculation corresponding to a process that tunnels between two degenerate vacua, for example, $\ket{\theta-\frac{\theta_\rho}{\sqrt 2 m}=0}$ and $\ket{\theta-\frac{\theta_\rho}{\sqrt 2 m}=\frac{\pi}{m}}$. This process can be intuitively understood as the process of a $\frac{hc}{2e}$ vortex encircling just the nanowire region, shown as process (b) in  Fig.~\ref{fig:splitting}. The instanton corresponding to such an event is homogeneous in space resulting in the splitting $\delta E_1\sim \exp\left(-\frac{4\sqrt{K}}{\pi m}\frac{l}{\xi}\right)$ where $l$ is the length of the nanowire segment as shown in Fig. \ref{fig:splitting} and $\xi=v/\Delta_P$. Such a process essentially shifts $J$ by $1$, and can be incorporated into the low energy Hamiltonian as $H_{S1}=\delta E_b\ket{J}\bra{J+1}+h.c.$.

\begin{figure}[t]
	\centering
	\includegraphics[width=0.9\columnwidth]{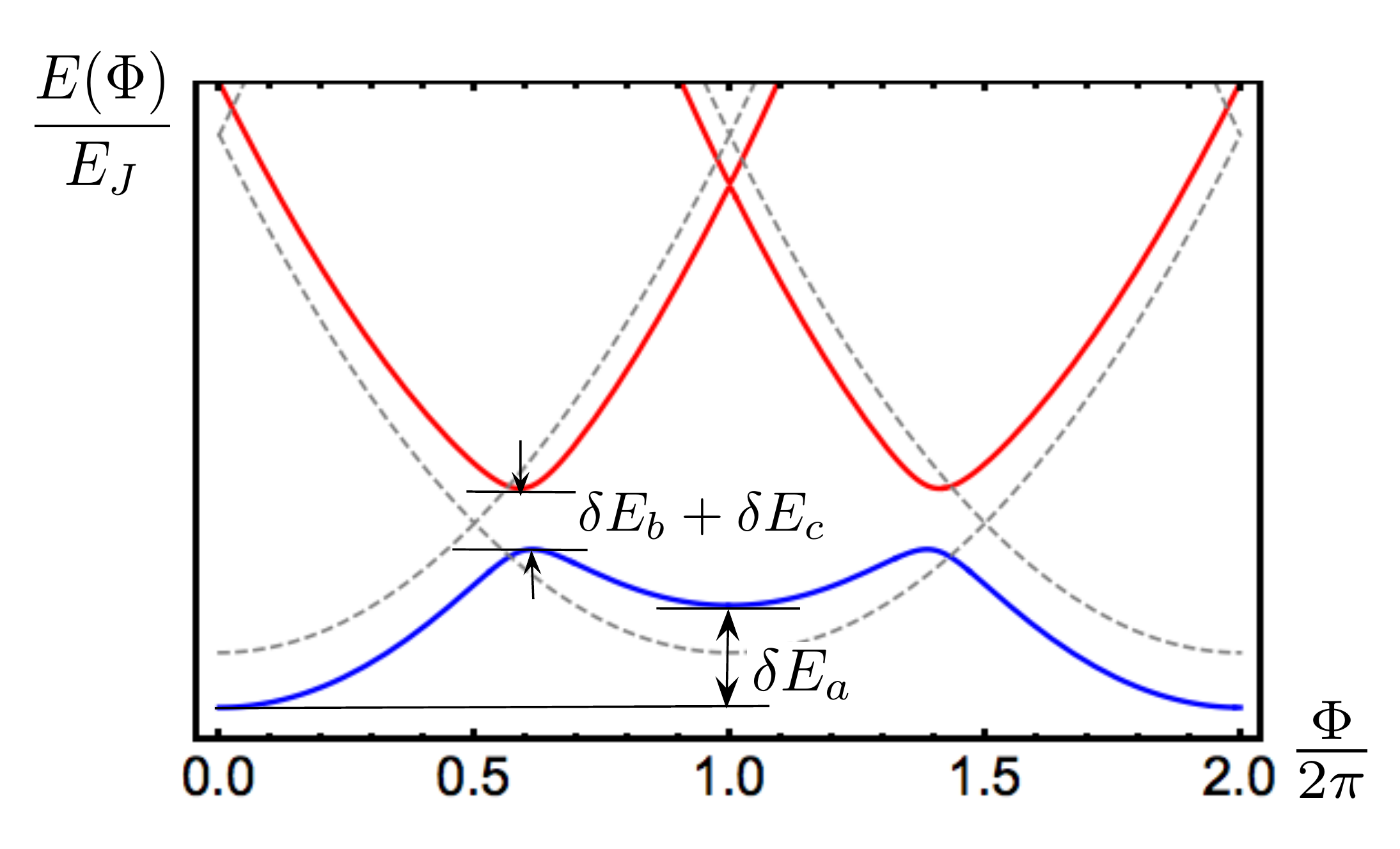}
	\caption{Ground-state energy dependance on magnetic flux $\Phi$ through the loop for $m=1$. Instanton tunneling processes $\delta E_b$ and $\delta E_c$ lead to the hybridization between different topological sectors whereas $\delta E_a$ process results in even-odd effect.}
	\label{fig:splitting2}
\end{figure}

We now discuss electron backscattering effects which generate process (c). Consider, for example, electron backscattering in the QSC given by the Hamiltonian
\begin{align}
H_{\rm imp}=\sum_i v_i \int dx\, \delta(x-x_i) \cos \sqrt 2 \varphi_\rho (x).
\end{align}
One can see that $\sqrt{2}\varphi_\rho$ creates a kink of $\pi$ in the dual field $\frac{\theta_\rho}{\sqrt{2}}$, and therefore can be thought as a phase slip event. Indeed, a $2\pi$ phase slip created in the phase field $\sqrt{2}\theta_\rho$ can be intuitively understood as an $\frac{hc}{2e}$ vortex tunneling across the QSC. Such vortex measures the fermion parity in the topological nanowire plus an underlying QSC~\footnote{Notice that the vortex actually circles not only the nanowire, but also the part of the QSC that are connected to the nanowire by pair tunneling.}, causing a splitting of the ground state degeneracy.  Since vortex actually measures the charge of the encircled region, see Fig.~\ref{fig:splitting}, such a  process is associated with the effective charging energy of the enclosed region involving both the nanowire as well as the QSC. We note that impurity scattering in the pair tunneling region is suppressed. Therefore, we consider the effect of impurities outside of this region. The details of the splitting calculation are presented in Appendix~\ref{app:splitting}. Here we simply discuss our main results. To simplify the instanton calculation, we consider the system shown in Fig.~\ref{fig:splitting} with two impurities at the positions $x_1$ and $x_2$. Impurity backscattering in the QSC leads to the following splitting energy
\begin{align}
H_c=\delta E_c\ket{J}\bra{J+1}+\text{h.c.}
\end{align}
where the energy $\delta E_c$ scales as a power-law of the system size. This is an important consequence of quantum fluctuations: topological ground-state degeneracy does not scale exponentially as in the case of a bulk superconductor but rather as a power-law. In the limit $K_\rho \gg K$, the splitting energy becomes $\delta E_c \propto v_1 v_2 |x_1-x_2|^{1-K_\rho}$. If we replace an impurity at $x_2$, for example, by the hard-wall boundary, the splitting energy is given by $\delta E_c \propto v_1 |x_1-x_2|^{1-K_\rho/2}$ which is simply determined by the scaling dimension of the $\cos \sqrt 2 \varphi_\rho$ operator, see Ref. \onlinecite{Fidkowski2011}.

Next, we consider impurity scattering in the nanowire described by the following Hamiltonian:
\begin{align}
H^{\rm NW}_{\rm imp}=\sum_i v_i \int dx\, \delta(x-x_i) \cos 2m \varphi (x)
\end{align}
One can notice that an operator $e^{i2m\varphi(x)}$ acting on the ground state, characterized by the field $\theta_-=\theta-\theta_\rho/\sqrt 2 m$, shifts $\theta_-$ by $2\pi m$ and, therefore, does not induce any transitions between degenerate ground states.

\begin{figure}
	\centering
	\includegraphics[width=1\columnwidth]{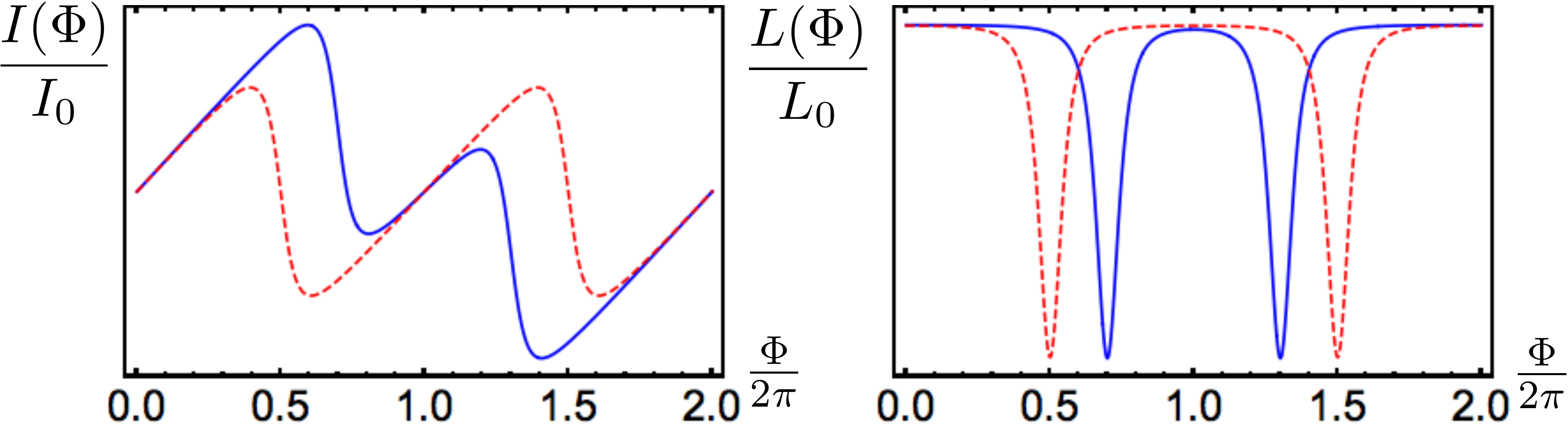}
	\caption{Schematic plot of the supercurrent and inductance of the ring as a function of the magnetic flux $\Phi$. Observables such as supercurrent and inductance develop $4\pi$ periodicity once $\delta E_a$ is finite. Red (dashed) and blue(solid) lines correspond to $\delta E_a=0$ and $\delta E_a/E_J=0.1$, respectively.}
	\label{fig:current}
\end{figure}

In summary, we have derived low energy theory for the superconducting loop structure shown in Fig.~\ref{fig:setup}. The corresponding Hamiltonian can be written as
\begin{align}
		H_\text{eff}&=\left(E_J\left(J+\frac{\Phi}{2\pi}\right)^2 +\delta E_a(-1)^J\right)\ket{J}\bra{J}\nonumber\\
		&+(\delta E_b+\delta E_c)\ket{J}\bra{J+1}+\text{h.c.},
  \label{eq:effHamiltonian}
\end{align}
where $J=2mk+n_J$ and $E_J=\frac{2 f_0^2}{\pi m^2 L_{\rm loop}}$. The spectrum of the system is shown in Fig.~\ref{fig:splitting2}. One can see that there are two types of non-commuting splitting terms. The processes described by $\delta E_b$ and $\delta E_c$ splitting energies result in the hybridization between different $J$ and $J+1$ sectors and, thus, open a gap at the avoided level crossings. This is not surprising since fractional excitations have an associated charge $e/m$, and thus charging energy inducing vortex tunneling can distinguish between different topological sectors and couple them. On the other hand, we have a process causing the splitting energy $\delta E_a$ which lifts the degeneracy between even and odd $J$-sectors. In particular, in $m=1$ case the splitting energy $\delta E_a$ distinguishes between even and odd parity sectors and restores $4\pi$ periodicity of the fractional Josephson effect, see Fig.~\ref{fig:splitting2}. The results obtained via the bosonization analysis of a microscopic Hamiltonian agrees well at the qualitative level with the simple effective Hamiltonian~\eqref{eqn:bcs} subject to the total fermion parity constraint, see Sec.\ref{sec:qual}.

\section{Conclusions}

In this paper we study Josephson effect in mesoscopic superconducting structures and focus on the contribution to the supercurrent due to the presence of Majorana zero-energy modes or more exotic parafermionic modes. In the former case, coherent single electron tunneling across the Josephson junction is allowed due to ground state degeneracy and results in the $4\pi$-periodic (fractional) contribution to the Josephson current. Previously, it was believed that fractional Josephson effect may only be accessed in {\it ac} measurements~\cite{Kwon} which might be quite challenging in realistic experimental settings~\cite{Badiane2013}. Here we show that one might be able to detect this anomalous periodicity with magnetic flux in mesoscopic superconducting rings using {\it dc} measurements in which the flux through the ring is changed adiabatically, and, therefore, to avoid the aforementioned issues.

The system we consider here consists of a quasi-one-dimensional superconductor (no long-range order) coupled to a nanowire or an edge carrying $e/m$ fractional charge excitations with $m$ being an odd integer. Using bosonization technique and instanton analysis, we derive an effective low-energy Hamiltonian for the system by taking into account instanton events which couples different topological sectors, see Eq. ~\eqref{eq:effHamiltonian}. We show that quantum phase fluctuations lead to a power-law dependence of the splitting energy with the distance. This is to be contrasted with the splitting energy in long-range-ordered superconductors which scales exponentially with the system size. We calculate the periodicity of the supercurrent on magnetic flux piercing the superconducting loop and show that, by properly designing mesoscopic ring, one should be able to measure $4\pi$-periodic component of the supercurrent in {\it dc} experiments. We believe that our proposal is within the experimental reach and provide here detailed recipe for measuring fractional {\it dc} Josephson effect in mesoscopic systems: \\
\begin{itemize}
  \item fabricate a mesoscopic ring with approximate dimensions of $100$nm in diameter and $10\mu$m circumference such that the charging energy of the ring is much larger than the dilution fridge temperature. The circumference of the ring should be much larger than the superconducting coherence length which is estimated to be of the order of $100$nm  (i.e. $L_{12}, L_{34} \gg \xi$), see Fig.\ref{fig:setup}.
  \item bring the outer ends of a semiconducting nanowire within the superconducting coherence length $\xi$ so that there is an appreciable splitting energy $\delta E_{14}$ (for $L_{14}\lesssim \xi$, the splitting energy $\delta E_{14}$ can be of the order of p-wave gap).
  \item measure the supercurrent in the ring using, for example, SQUID~\cite{Bluhm09} or torque~\cite{Harris2009, Jang2011} magnetometry technique. We estimated the supercurrent in the ring to be of the order of $10n$A which should be within experimental reach.
  \item the semiconductor nanowire should be designed in such a way so that one can drive topological quantum phase transition with an in-plane magnetic field (for example, $U$-shape nanowire). In the trivial phase (e.g. at zero magnetic field $B=0$), the ground state energy exhibits $2\pi$-periodic dependence on the external flux whereas in the non-trivial phase the flux periodicity should be $4\pi$.
\end{itemize}

\acknowledgements

We thank J. Alicea, M. Barkeshli, D. Clarke, L. Fidkowski, Dong Liu and C. Nayak for stimulating discussions. RL wishes to acknowledge the hospitality of the Aspen Center for Physics and support under NSF Grant \#1066293.

\appendix
\onecolumngrid
\section{Degeneracy splitting due to quantum phase slips}\label{app:splitting}
In this Appendix, we calculate the topological degeneracy splitting due to quantum phase slips. To be concrete, we consider quantum phase slips generated by impurity backscattering. For simplicity, we consider the geometry shown in   Fig.\ref{fig:splitting} where the impurities are located at the positions $x_1$ and $x_2$, just outside the pair-hopping region. For simplicity, we consider $m=1$ case. The corresponding Hamiltonian then reads
\begin{align}
H_{\rm imp}=v_1 \cos \sqrt 2 \varphi_\rho (x_1)+v_2 \cos \sqrt 2 \varphi_\rho (x_2).
\end{align}
We now compute the amplitude $M$ for the instanton tunneling between the minima of $\cos(2\theta-\sqrt 2 \theta_\rho)$ in the interval $x_1<x<x_2$ :
\begin{equation}
M=\left\langle \theta-\frac{\theta_\rho}{\sqrt{2}}=\pi\Big|e^{-HT}\Big|\theta-\frac{\theta_\rho}{\sqrt{2}}=0\right\rangle_{T\rightarrow \infty},
	\label{}
\end{equation}
where $H$ is the total Hamiltonian including $H_{\rm imp}$. This amplitude can be calculated using a path integral with the appropriate boundary conditions:
\begin{align}\label{eq:part_constraint}
\mathcal{Z}=\sum_{k_1, k_2 \in \mathbb{Z}}\int\mathcal{D}\varphi\mathcal{D}\theta\, \delta\left(\theta_-(x,0)-\pi(2k_1+1)\right)\delta\left(\theta_-(x,T)-2 \pi k_2\right) e^{-\int\di\tau\di x (\mathcal{L}+\mathcal{L}_{\rm imp})}
\end{align}
where $\theta_-(x,\tau)=\theta(x,\tau)-\frac{\theta_\rho(x,\tau)}{\sqrt{2}}$ and
\begin{align}
\mathcal{L}&=\frac{i}{\pi}\partial_x\varphi_\rho\partial_\tau \theta_\rho+\frac{i}{\pi}\partial_x\varphi\partial_\tau \theta+\frac{v}{2\pi}[K^{-1}(\partial_x\varphi)^2+K(\partial_x\theta)^2]+\frac{v_\rho}{2\pi}[K_\rho^{-1}(\partial_x\varphi_\rho)^2+K_\rho(\partial_x\theta_\rho)^2]\nonumber\\
	&-\frac{\Delta_p}{2\pi a}\Theta(x-x_1)\Theta(x_2-x)\cos [2\theta(x)-\sqrt{2}\theta_\rho(x)],\\
\mathcal{L}_{\rm imp}&= [v_1\delta(x-x_1^-)+v_2\delta(x-x_2^+)]\cos \sqrt{2}\varphi_\rho(x,\tau).
\end{align}
Here the pairing tunneling term is non-zero in the interval $x_1<x<x_2$ which defines the topological wire segment. In the limit $K_\rho >2$ when backscattering is irrelevant, we can calculate partition function perturbatively in $v_i$
\begin{equation}
	\mathcal{Z}=\int\mathcal{D}\varphi\mathcal{D}'{\theta}\, e^{-\int\di\tau\di x \mathcal{L}}=\int\mathcal{D}\varphi\mathcal{D}\theta\, \left(1+\frac{v_1 v_2 }{2}\int\di\tau_1\di\tau_2\,\cos\sqrt{2}\varphi_\rho(x_1^-,\tau_1)\cos\sqrt{2}\varphi_\rho(x_2^+,\tau_2)+\cdots\right)e^{-\int\di \tau\di x\mathcal{L}_0}.
\end{equation}
Here $\mathcal{D}'{\theta}$ denotes integration with the $\delta$ function constraints, see Eq.\eqref{eq:part_constraint}.
Using the identity
\begin{equation}
	\cos\sqrt{2}\varphi_\rho(x_1^-,\tau_1)\cos\sqrt{2}\varphi_\rho(x_2^+,\tau_2)=\frac{1}{4}\sum_{s_1,s_2=\pm 1}e^{i\sqrt{2}s_1\varphi_\rho(x_1^-)}e^{i\sqrt{2}s_2\varphi_\rho(x_2^+)},
	\label{}
\end{equation}
one finds that the effective action for $\varphi_\rho$ is given by
 \begin{equation}
	 \mathcal{S}[\varphi_\rho]=\int\di x\di \tau\left[\frac{i}{\pi}\partial_x\varphi_\rho\partial_\tau \theta_\rho+\frac{v_\rho}{2\pi}K_\rho^{-1}(\partial_x\varphi)^2\right]+ i\sqrt{2}\left[s_1\varphi_\rho(x_1^-,\tau_1)+s_2\varphi_\rho(x_2^+,\tau_2)\right].
\end{equation}
After integrating out $\varphi$,
the effective action becomes
 \begin{equation}
	 \begin{split}
		 \mathcal{S}[\varphi_\rho]&=\int\frac{\di^2\vec{q}}{(2\pi)^2}\left[\frac{i}{\pi}\omega k\varphi_\rho(\vec{q}) \theta_\rho(-\vec{q})+\frac{v_\rho k^2}{2\pi K_\rho}\varphi_\rho(\vec{q})\varphi_\rho(-\vec{q})+ \sqrt{2}i(s_1e^{i\vec{q}\cdot\vec{x}_1}+s_2e^{i\vec{q}\cdot\vec{x}_2})\varphi_\rho(\vec{q})\right]\\
		 &=\int\frac{\di^2\vec{q}}{(2\pi)^2}\left[\frac{vk^2}{2\pi K_\rho}\varphi_\rho(\vec{q})\varphi_\rho(-\vec{q})+i\left(\frac{\omega k}{\pi}\theta_\rho(-\vec{q})+\sqrt{2}s_1e^{i\vec{q}\cdot\vec{x}_1}+\sqrt{2}s_2e^{i\vec{q}\cdot \vec{x}_2}\right)\varphi_\rho(\vec{q})\right],
	 \end{split}
\end{equation}
where $\vec{q}=(k,w), \vec{x}=(x,\tau)$ and the inner product $\vec{q}\cdot\vec{x}=kx-\omega\tau$. We can now integrate out $\varphi_\rho$ to find
\begin{equation}
	 \begin{split}
		 \mathcal{S}[\theta_\rho]&=-\int\frac{\di^2\vec{q}}{(2\pi)^2}\frac{\pi K_\rho}{2v_\rho k^2}\left|\frac{\omega k}{\pi}\theta_\rho(\vec{q})+\sqrt{2}s_1e^{-i\vec{q}\cdot\vec{x}_1}+\sqrt{2}s_2e^{-i\vec{q}\cdot \vec{x}_2}\right|^2\\
		 &=\frac{K_\rho}{2\pi v_\rho }\int\frac{\di^2\vec{q}}{(2\pi)^2}\left|\omega\theta_\rho(\vec{q})+\frac{\sqrt{2}\pi s_1}{k}e^{-i\vec{q}\cdot\vec{x}_1}+\frac{\sqrt{2}\pi s_2}{k}e^{-i\vec{q}\cdot \vec{x}_2}\right|^2\\
		 &=\frac{K_\rho}{2\pi v_\rho }\int\di\tau\di x\,\left[\partial_\tau \theta_\rho+\sqrt{2}\pi s_1\Theta(x-x_1)\delta(\tau-\tau_1)+\sqrt{2}\pi s_2\Theta(x-x_2)\delta(\tau-\tau_2)\right]^2
	 \end{split}
\end{equation}

Next, we integrate out $\varphi$ and obtain the following effective action:
\begin{multline}
	\mathcal{L}_{s_1s_2}[\theta_\rho,\theta]=\frac{K_\rho}{2\pi}\left[\frac{1}{v_\rho}\left(\partial_\tau \theta_\rho+\sqrt{2}\pi s_1\Theta(x_1-x)\delta(\tau-\tau_1)+\sqrt{2}\pi s_2\Theta(x_2-x)\delta(\tau-\tau_2)\right)^2 +v_\rho(\partial_x\theta_\rho)^2\right]\\
	+\frac{K}{2\pi}\left[\frac{1}{v}(\partial_\tau\theta)^2+v(\partial_x\theta)^2\right]
	-\frac{\Delta_p}{2\pi a}\Theta(x-x_1)\Theta(x_2-x)\cos \big[2\theta(x)-\sqrt{2}\theta_\rho(x)\big].
	\label{}
\end{multline}
Combining all the terms together, the partition function now reads
\begin{equation} \mathcal{Z}=\int\mathcal{D}\theta_\rho\mathcal{D}'{\theta}\left(e^{-\mathcal{S}_0}+\frac{v_1 v_2}{8}\int d \tau_1 d \tau_2 \sum_{s_1s_2}e^{-\mathcal{S}_{s_1s_2}}+\cdots\right).
	\label{}
\end{equation}
The calculation of the first term $\int \mathcal{D}'\theta\mathcal{D}\theta_\rho e^{-\mathcal{S}_0}$ reproduces the splitting energy $\delta E_1$, see Sec.\ref{sec:splitting}. We will focus here on the second term and calculate the contribution of the classical field configuration minimizing the action $\mathcal{S}_{s_1s_2}$. One can notice that the the quadratic action of $\theta_\rho$ contains $\delta$ function in $\tau$. Therefore, in order to get a finite action $\theta_\rho$ field must have a discontinuity at $\tau_1$ and $\tau_2$. Indeed, let us write $\theta_\rho$ as
\begin{equation}
	\theta_\rho=\tilde{\theta}_\rho-\mathcal{A}(x), \mathcal{A}(x)=\sqrt{2}\pi [s_1\Theta(x_1-x)\Theta(\tau-\tau_1)+s_2\Theta(x_2-x)\Theta(\tau-\tau_2)],
	\label{}
\end{equation}
where $\tilde{\theta}_\rho (x, \tau)$ is a now smooth field as far as time dependence is concerned. Notice that in doing so we have introduced a jump in the spatial profile of $\theta_\rho$ at $x_1$ and $x_2$. The discontinuity at these points has to be carefully taken into account by considering an inhomogeneous problem since the pairing field also has jumps at $x_1$ and $x_2$. However, in the limit when the length of the topological region is large, the bulk energy gives dominant contribution and thus the boundary effects can be ignored.

Next, we rewrite the action using the new fields:
\begin{equation}
	\begin{split}
		\mathcal{L}_{s_1s_2}[\theta_\rho,\theta]=&\frac{K_\rho}{2\pi}\left[\frac{1}{v_\rho}(\partial_\tau \tilde{\theta}_\rho)^2 +v_\rho(\partial_x\tilde{\theta}_\rho-\partial_x\mathcal{A})^2\right]\\
	&+\frac{K}{2\pi}\left[\frac{1}{v}(\partial_\tau\theta)^2+v(\partial_x\theta)^2\right]
	-\frac{\Delta_p}{2\pi a}\Theta(x-x_1)\Theta(x_2-x)\cos \big[2\theta(x)-\sqrt{2}\tilde{\theta}_\rho(x)\big].	
	\end{split}
	\label{}
\end{equation}
In the domain $x_1 <x<x_2$, the combination $\theta-\tilde{\theta}_\rho/\sqrt{2}$ is pinned and one can use the relation $\partial_{x/\tau}\theta=\partial_{x/\tau}\tilde{\theta}_\rho/\sqrt{2}$ to simplify the calculation. Thus, within this space of field configurations, the corresponding partition function can be evaluated exactly
\begin{align}
	\mathcal{Z}_{v_1 v_2}&=\frac{v_1 v_2}{4} \int d \tau_1 d \tau_2 \int\mathcal{D}\tilde{\theta}_\rho\, e^{-\mathcal{S_{\rm eff}}}\\
\mathcal{S}_{\rm eff}[\theta]&=\int d\tau dx \frac{1}{2\pi}\left[\left(\frac{K_\rho}{v_\rho}+\frac{K}{2v}\right)(\partial_\tau \tilde{\theta}_\rho)^2 +v_\rho K_\rho(\partial_x\tilde{\theta}_\rho-\partial_x\mathcal{A})^2 + \frac{vK}{2}(\partial_x \tilde{\theta}_\rho)^2\right]
	\label{}
\end{align}
We can now evaluate the path integral and calculate the dependence of this term on $L=x_2-x_1$. Let us look closer at the term $(\partial_x\tilde{\theta}-\mathcal{A})^2$:
\begin{equation}
	\int\di x\di\tau\,(\partial_x\tilde{\theta}_\rho-\partial_x\mathcal{A})^2\!=\!\int\di\tau\!\! \int\di x\ \,\left[(\partial_x\tilde{\theta}_\rho)^2-2\sqrt{2}\pi(\partial_x\tilde{\theta}_\rho)(x_1,\tau)\Theta(\tau\!-\!\tau_1)+2\sqrt{2}\pi(\partial_x\tilde{\theta}_\rho)(x_2,\tau)\Theta(\tau\!-\!\tau_2)+(\partial_x\mathcal{A})^2\right]
	\label{eq:artifact}
\end{equation}
For our purpose, it is sufficient to consider large $L$ limit and neglect the boundary conditions for $\tilde{\theta}$ which does not affect the scaling of the action with $L$. Therefore, we neglect the divergent boundary term $(\partial_x\mathcal{A})^2$ in Eq.\eqref{eq:artifact} which is simply an artifact of our approximation. The effective partition function in the momentum space becomes
\begin{equation}
	\mathcal{S}_\text{eff}=\int\frac{\di k\di \omega}{(2\pi)^2}\left\{\frac{1}{2\pi}\left[\left( \frac{K_\rho}{v_\rho}+\frac{K}{2v} \right)\omega^2+\left(v_\rho K_\rho+\frac{vK}{2}\right)k^2\right]|\tilde{\theta}_\rho(\vec{q})|^2 + \sqrt{2} v_\rho K_\rho \frac{k}{\omega}(s_1e^{i\vec{q}\cdot\vec{x}_2}\!+\!s_2e^{i\vec{q}\cdot\vec{x}_1})\tilde{\theta}_\rho(\vec{q})\right\}.
	\label{}
\end{equation}
It is clear that the correlation function is vanishing for $s_1=s_2$. Therefore, in the following we set $s_1=1, s_2=-1$. After integrating out $\tilde{\theta}$ field, one finds
\begin{align}
	\int\mathcal{D}\tilde{\theta}_\rho\, e^{-\mathcal{S}_\text{eff}}\!&\!=\!\exp\left(-\int\frac{\di k\di \omega}{4\pi} \frac{v_\rho^2v_+ K_\rho^2k^2}{K_+\omega^2(\omega^2+v_+^2 k^2)}|e^{i\vec{q}\cdot\vec{x}_2}-e^{i\vec{q}\cdot\vec{x}_1}|^2\right)
	\!=\!C_1\exp\left(-\frac{v_\rho^2 K_\rho^2}{2v_+^2K_+ }\ln\left[\frac{v_+^2 |\tau_{1}\!-\!\tau_2|^2\!+\!|x_1\!-\!x_2|^2}{a^2}\right]\right),
	\label{}
\end{align}
where $a$ is a UV cutoff and $C_1$ is the numerical prefactor. Finally, the contribution to the partition function reads
\begin{equation}
	\begin{split}
	\mathcal{Z}_{s_1s_2}= \frac{v_1 v_2 C_1}{8} \int\di\tau_1\di\tau_2\,\left(\frac{a^2}{v_+^2 |\tau_{1}-\tau_2|^2+|x_1-x_2|^2} \right)^{\frac{v_\rho^2 K_\rho^2}{2v_+^2K_+ }}
	&\sim \frac{v_1 v_2 T a}{v_+} \frac{1}{|x_1-x_2|^{\frac{v_\rho^2 K_\rho^2}{v_+^2K_+ }-1}}
	\end{split}
	\label{}
\end{equation}
Following standard calculation, see Ref.~[\onlinecite{Coleman85}], in order to obtain the energy splitting, we need to take into account multiple instanton processes corresponding to multiple insertions of jumps of $\tilde{\theta}_\rho$. In the end, one finds that the energy splitting is given by
\begin{equation}
	\delta E\sim \frac{v_1 v_2 a}{v_+} \frac{1}{|x_1-x_2|^{\frac{v_\rho^2 K_\rho^2}{v_+^2K_+ }-1}},
	\label{}
\end{equation}
and is power-law dependent on the system size. In the limit $K_\rho \gg K$, $v_+ \approx v_\rho$ and $K_+\approx K_\rho$, so the splitting energy becomes $\delta E \propto L^{1-K_\rho}$. Note that the boundary of a QSC can be represented as a strong impurity, say, at $x_2$ which effectively cuts off the superconductor at this point. In this case, the fluctuations of the phase $\varphi_\rho$ at $x_2$ are suppressed, and the splitting is given by scaling dimension of $\cos \sqrt 2 \varphi_\rho(x_1)$~\cite{Fidkowski2011}, i.e. $\delta E \propto |x_1-x_2|^{1-K_\rho/2}$ at $K_\rho \gg K$.

\twocolumngrid

\bibliography{topological_wires}

\end{document}